\documentclass[draftclsnofoot, 12pt,onecolumn,oneside,compress]{IEEEtran}

\ifCLASSINFOpdf
\else
   \usepackage[dvips]{graphicx}
\fi
\usepackage{url}

\usepackage{graphicx,balance}
\usepackage{bm}
\usepackage{amssymb}
\usepackage{amsmath}
\usepackage{textcomp}
\usepackage{color}
\usepackage{epstopdf}
\usepackage{subfigure}
\usepackage{booktabs}
\usepackage{enumitem}
\usepackage{algorithm}
\usepackage{algorithmic}

\begin{document}

\title{Deep Learning Based Channel Covariance Matrix Estimation with User Location and Scene Images}

\author{Weihua Xu, Feifei Gao, Jianhua Zhang, Xiaoming Tao, and Ahmed Alkhateeb
\thanks{Manuscript received April 10, 2021; revised July 15, 2021; accepted
August 17, 2021. This work was supported in part by National Key Research and Development Program of China (2018AAA0102401), by the National Natural Science Foundation of China under Grant \{61831013, 61925102\}. \emph{(Corresponding author: Feifei Gao.)}}
\thanks{W. Xu and F. Gao are with Institute for Artificial Intelligence, Tsinghua University (THUAI), Beijing National Research Center for Information
Science and Technology (BNRist), Department of Automation, Tsinghua
University, Beijing, P.R. China, 100084 (email: xwh19@mails.tsinghua.edu.cn,
feifeigao@ieee.org).}
\thanks{J. Zhang is with the State Key Laboratory of Networking and Switching Technology, Beijing University of Posts
and Telecommunications, Beijing 100876, China (e-mail: jhzhang@bupt.edu.cn).}
\thanks{X. Tao is with the Department of Electronic Engineering, Tsinghua University, Beijing, P.R. China, 100084 (email: taoxm@tsinghua.edu.cn).}
\thanks{A. Alkhateeb is with the Department of Electrical and Computer Engineering, University of Texas at Austin, Austin, TX 78712-1687 USA (e-mail:
alkhateeb@asu.edu).}}

\maketitle
\vspace{-7mm}
\begin{abstract}
Channel covariance matrix (CCM) is one critical parameter for designing the communications systems. In this paper, a novel framework of the deep learning (DL) based CCM estimation is proposed that exploits the perception of the transmission environment without any channel sample or the  pilot signals. Specifically, as CCM is affected by the user's movement, we design a deep neural network (DNN) to predict CCM from user location and user speed, and the corresponding estimation method is named as ULCCME. A location denoising method is further developed to reduce the positioning error and improve the robustness of ULCCME. For cases when user location information is not available, we propose an interesting way that uses the environmental 3D images to predict the CCM, and the corresponding estimation method is named as SICCME. Simulation results show that both the proposed methods are effective and will benefit the subsequent channel estimation.
\end{abstract}


\IEEEpeerreviewmaketitle

\section{Introduction}

\IEEEPARstart{C}{hannel} covariance matrix (CCM) is an essential information to enhance the performance of channel estimation and beamforming \cite{Xie}, especially in massive MIMO system \cite{Roh}. However, traditional CCM estimation needs long training signals to build CCM which is vulnerable to be inaccurate or outdated. Therefore, acquiring CCM with low overhead has received much attention in recent years \cite{Liang}-\cite{wang1}.

In \cite{Liang}-\cite{Decu}, the authors resorted to the uplink CCM to predict the  downlink CCM for the frequency division duplex (FDD) massive MIMO system without the need of downlink channel estimation. However, these methods still need huge overhead during uplink channel estimation especially when the base station (BS) contains large number of antennas. Based on the sparse characteristics of millimeter wave channels, the authors of \cite{Ahmed} applied compressive sensing to estimate CCM with low training overhead. The authors of \cite{Giordani} translated the CCM  from sub-6 GHz to the mmWave band by compressed signal recovery theorem and thus reduced the overhead of CCM estimation at mmWave band. Though the methods of \cite{Ahmed} and \cite{Giordani} can improve the efficiency of CCM estimation, they highly rely on the assumption of channel sparsity. In \cite{wang1}, a deep learning (DL) based CCM estimation is designed by utilizing rich channel information contained in the uplink signals at multiple base stations (BSs). Although \cite{wang1} has demonstrated that DL method can significantly reduce the training overhead compared with compressive sensing methods, it requires coordination among multiple BSs and is hard to implement in practice. Nevertheless, \cite{wang1} brought the insight that the powerful DL tools can be potentially used to improve CCM estimation and deserves further exploration.\\
\indent
It was recently found that the side information of communication environments, such as user locations \cite{zhou}-\cite{Aviles}, scene point clouds \cite{xing2}-\cite{Xu}, scene RGB images \cite{Ahmed1}-\cite{Ahmed3}, can reflect rich channel characteristics with the aid of strong modeling ability of machine learning \cite{HHe1}-\cite{HHe3}. Specifically, most system parameters of channel matrix can be regarded as functions of the environment information, like objects' distributions, shapes, materials, etc \cite{Zhang1}-\cite{Zhang3}. These functions are difficult to express mathematically or model accurately, but can be well represented by the machine learning models and can be trained by the real sample data in an offline manner. Actually, side information has already been exploited to enhance the efficiency of beam alignment \cite{zhou}-\cite{Ahmed3}. It then motivates us to design a highly efficient and more general approach for CCM estimation from the environmental information and thus reduce or even remove the requirement of training sequences.\\
\indent
In this paper, we present a novel framework to utilize the side information such as user location, user speed, and the scene images surrounding the user, etc., to estimate CCM in a \emph{pilot-free} manner. The main contributions of this paper are listed as follows:
\begin{itemize}
  \item[1)]
  User location based CCM estimation (ULCCME): We propose to first estimate a rough user moving region that contains all possible user locations at a certain moment. Then a location based CCM estimation DNN, called LCNET, is designed to learn the mapping from the user's starting location and the speed to CCM of the corresponding user moving region. The well-trained DNN can be utilized to predict  CCM for estimating the channel at any concerned moment.
  \item[2)]
  Deep learning based location denoising: As the user locations are hard to be estimated accurately, we design a location denoising algorithm to reduce such error. We first  train a location estimation DNN, called LENET, to learn the mapping from the user channel to the user location. Then, we utilize the user speed, the error distribution of LENET and the noise distribution of the uploaded location to correct the uploaded user location through the Kalman framework.
  \item[3)]
  Scene image based CCM estimation (SICCME): When the user location is unavailable due to the strong noise corruption or the protection of the user's privacy, we propose to utilize the scene images of the user's surrounding environment to replace the user location as the input feature of DNN, and the so obtained image based DNN is called as ICNET.
\end{itemize}

\indent
This paper is organized as follows. Section II introduces the signal model, and illustrates the existence of the mapping between user moving trajectory and channel covariance. Section III proposes the ULCCME method and the location denosing method, while Section IV presents the SICCME method. Section V describes the performance metric, the simulation setup, the data set generation, and then provides numerous simulation results and fruitful discussions. Finally, the conclusions are drawn in Section VI.\\
\indent
Notation: $\bm{A}$ is a matrix; $\bm{a}$ is a vector; a is a scalar; $[\bm{a}]_{i}$ is the $i$th element of $\bm{a}$; $\bm{a}_{i:j}$ is the column vector $[[\bm{a}]_{i},[\bm{a}]_{i+1},\cdots,[\bm{a}]_{j}]^{\mathrm{T}}$; $[\bm{A}]_{i,j}$ is the element of the $i$th row and $j$th column in $\bm{A}$; $[\bm{A}]_{i,:}$ and $[\bm{A}]_{:,j}$ are the $i$th row and the $j$th column of $\bm{A}$ respectively; $\bm{I}_r$ is the identity matrix with rank $r$ and $\mathcal{N}(\bm{m}_{\mathrm{g}}, \bm{R}_{\mathrm{g}})/\mathcal{CN}(\bm{m}_{\mathrm{g}}, \bm{R}_{\mathrm{g}})$ is the real/complex Gaussian
random distribution with mean $\bm{m}_{\mathrm{g}}$ and covariance $\bm{R}_{\mathrm{g}}$; $\mathrm{E}\{\ \cdot\ \}$ is the expectation operator.

\section{Problem Formulation}
\subsection{Signal Model}
We consider downlink communications with one BS and one user. BS is equipped with a uniform planar array (UPA) of $N_\mathrm{B}=N_\mathrm{ele}^{\mathrm{B}}\times N_\mathrm{az}^{\mathrm{B}}$ antennas, and the user is equipped with one single antenna. The signal received at the user can be expressed as
\begin{equation}
y=\bm{h}^{\mathrm{H}}\bm{s}+n,
\end{equation}
where $\bm{h}\in \mathbb{C}^{N_\mathrm{B}\times 1}$ is the downlink channel vector, $\bm{\mathrm{s}}\in \mathbb{C}^{N_\mathrm{B}\times 1}$ is the transmit signal at the BS, and $n \in \mathcal{CN}(0,\sigma^2)$ is the Gaussian noise.

We adopt the widely used geometric channel model \cite{Adm}
\begin{equation}
  \bm{h}=\sum_{l=1}^{L}\alpha_l\bm{a}_\mathrm{t}(\theta^t_l,\phi^t_l),
\end{equation}
where $\alpha_l$ is the complex gain of the $l$th path, $\theta^\mathrm{t}_l$ and $\phi^\mathrm{t}_l$ are the elevation and azimuth of the $l$th path's angle of departure, and $\bm{a}_\mathrm{t}(\theta,\phi) \in \mathbb{C}^{N_\mathrm{B}\times 1}$ is the complex steering vector of the transmit array.\\
\indent
The antenna spacing of UPA is set as $d$, and the steering vector $\bm{a}_{\mathrm{t}}(\theta,\phi)$ is given by
\begin{equation}
\bm{a}_{\mathrm{t}}(\theta,\phi)=\bm{a}_{\mathrm{az}}(\theta,\phi)\otimes\bm{a}_{\mathrm{ele}}(\theta),
\end{equation}
where
\begin{align}
\bm{a}_{\mathrm{ele}}(\theta)&=[1,e^{j\frac{2\pi d}{\lambda}\cos(\theta)},\cdots,e^{j(N^{\mathrm{B}}_{\mathrm{ele}}-1)\frac{2\pi d}{\lambda}\cos(\theta)}]^{\mathrm{T}},\\
\bm{a}_{\mathrm{az}}(\theta,\phi)&=[1,e^{j\frac{2\pi d}{\lambda}\sin(\theta)\sin(\phi)},\cdots,e^{j(N^{\mathrm{B}}_{\mathrm{az}}-1)\frac{2\pi d}{\lambda}\sin(\theta)\sin(\phi)}]^{\mathrm{T}},
\end{align}
$\lambda$ is the carrier wavelength and $\otimes$ represents the Kronecker product.

\subsection{The Mapping from User Location Distribution to CCM}
It is generally known that the channels depend on the scene objects between the BS and user, i.e., buildings, cars, trees, people, etc., as the propagation paths and attenuation of transmission signals can be completely determined by the environmental information and the BS/user locations when the carrier frequency is fixed. In fact, it can be assumed that there exists a one-to-one mapping between the user position and the channel \cite{Alrabeiah} at the fixed carrier frequency, i.e.,
\begin{equation}
\bm{p}_{\mathrm{user}}\mathop{\rightleftarrows}\limits_{\bm{M}^{-1}(\cdot)}^{\bm{M}(\cdot)}\bm{h}_{\mathrm{user}}, \label{gao:1}
\end{equation}
where $\bm{p}_{\mathrm{user}}\in \mathbb{R}^{3}$ is the 3D coordinates of the user location, $\bm{h}_{\mathrm{user}}$ is the channel corresponding to $\bm{p}_{\mathrm{user}}$, while $\bm{M}$ and $\bm{M}^{-1}$ represent the mapping function and the inverse mapping function respectively.

Unfortunately, since the accurate user location is generally an uncertain variable for BS (or user), measuring accurately the user location in real time, i.e., the localization problem, is very difficult \cite{Misra}-\cite{Zaidi}. Nevertheless, it would be easier for BS (or user) to obtain a probability distribution of the user location at a certain moment. Interestingly, the mapping function (\ref{gao:1}) also indicates that the user location distribution (ULD) can also reflect its channel distribution.

Let us denote the probability density function (PDF) of the user location at moment $t$ as $p_{\mathrm{u},t}(\bm{x})$, $\bm{x}\in\bm{C}$, where $\bm{x}\in\mathbb{R}^{3}$ is the 3D coordinates of the user location and $\bm{C}$ is a set that contains all possible values of $\bm{x}$. From (6), $p_{\mathrm{u},t}(\bm{x})$, $\bm{x}\in\bm{C}$ can determine a user's channel distribution and further determine the CCM at moment $t$. Specifically, CCM can be expressed as
\begin{equation}
\bm{\mathrm{Cov}}_{t}=\int_{\bm{C}} p_{\mathrm{u},t}(\bm{x})\bm{M}(\bm{x})\bm{M}^{\mathrm{H}}(\bm{x})\mathrm{d}\bm{x}.
\end{equation}
Then, the mapping from ULD to CCM can be expressed as
\begin{equation}
p_{\mathrm{u},t}(\bm{x}), \bm{x}\in\bm{C}\rightarrow\bm{\mathrm{Cov}}_{t}.
\end{equation}
Interestingly, (8) motivates us to utilize the ULD to directly obtain the CCM. Since the channels corresponding to closer user locations usually have similar path angles of departure, the pilot beam designed from the eigenvectors of the lager eigenvalues of the CCM are much likely to cover the direction of strong paths of the channel corresponding to $\bm{x}$, especially when the area $\bm{C}$ is sufficiently small, thereby offering high transmission gain to achieve high channel estimation accuracy. Hence, we expect the CCM from (8) could achieve better channel estimation performance than the estimated CCM from traditional approaches, when the latter is obtained from insufficient number of samples.

\textbf{Remark}: For simplicity, the mapping $\bm{p}_{\mathrm{user}} \rightarrow \bm{h}_{\mathrm{user}}$ shown in (6) is proposed for the static environment here, and the effect of dynamic objects on the channel is not considered. Practically, as the BS is usually placed in a high location and the sizes of ordinary dynamic objects, such as pedestrians and cars, are relatively small compared with the building sizes, the channel path distribution is basically determined by the surrounding buildings and the effect of dynamic objects on the strong channel paths is usually slight. Moreover, for the environment with dynamic objects, we can utilize the statistical channel $\mathrm{E}\{\bm{h}_{\mathrm{user}}\}$ to replace the instantaneous channel $\bm{h}_{\mathrm{user}}$ in the mapping (6), and thereby assume there exists the mapping from the user location to the statistical channel, i.e., $\bm{p}_{\mathrm{user}} \rightarrow \mathrm{E}\{\bm{h}_{\mathrm{user}}\}$. Then, the modified mapping: $\bm{p}_{\mathrm{user}} \rightarrow \mathrm{E}\{\bm{h}_{\mathrm{user}}\}$ can be used to obtain the channel covariance under dynamic environments by the equation (7), which still yields the mapping (8). Thus, the proposed mapping from ULD to CCM can be conveniently generalized for the dynamic environments.
\begin{figure*}[t]
\centering
\includegraphics[width=1\textwidth]{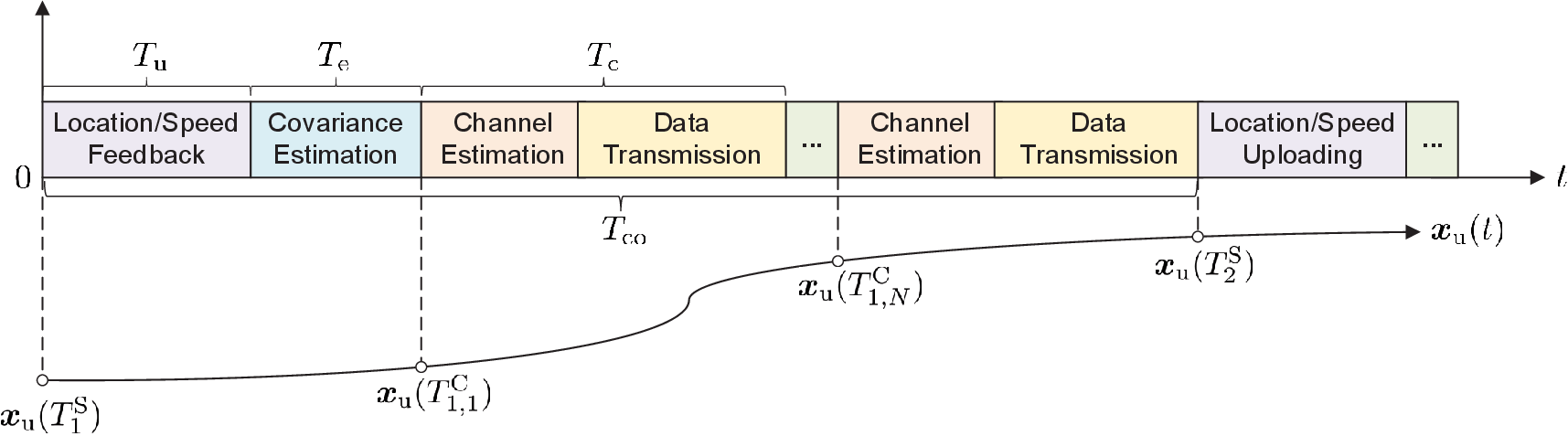}
\caption{The time domain diagram of the proposed user location based CCM estimation.}
\end{figure*}
\section{User Location Based CCM Estimation}
Fig.~1 shows the frame structure of the proposed ULCCME method. We consider a block fading scenario and assume the channel changes slowly during a time period $[\tau,\tau+T_{\mathrm{c}}]$ for any $\tau$, where $T_{\mathrm{c}}$ is the channel coherence time (CCT). Because the CCM is a slowly changing parameter compared to the instantaneous channel, we assume one CCM is used for multiple CCTs and define $T_{\mathrm{co}}$ as the covariance coherence time (COCT). For each COCT, the proposed frame structure includes three stages. In the first stage of length $T_u$, the user feeds back its location and speed to BS. In the second stage of length $T_e$, BS performs the CCM estimation utilizing the uploaded user location and speed. After CCM  estimation, data service will be provided during the third stage that is assumed to have length $NT_{\mathrm{c}}$. Within each CCT the linear minimum mean-square error (LMMSE) channel estimation can be performed with the aid of a small amount of pilot signals and the estimated CCM. Effective pilot signals can also be designed from the estimated CCM for each COCT.
\subsection{CCM  Estimation Method}
For the $k$th COCT, we assume the user location and speed at moment $T_{k}^{\mathrm{S}}=(k-1)T_{\mathrm{co}}$ are fed back to BS. While for the $q$th CCT of the $k$th COCT, the channel at moment $T_{k,q}^{\mathrm{C}}=(k-1)T_{\mathrm{co}}+T_{\mathrm{u}}+T_{\mathrm{e}}+(q-1)T_{\mathrm{c}}$ is estimated at the user and is fed back to BS. Define the user trajectory as $\bm{x}_{\mathrm{u}}(t)\in \mathbb{R}^{3\times 1}$ and express user speed as $v_{\mathrm{u}}(t)$, $t\in [0, +\infty)$. The channel corresponding to $\bm{x}_{\mathrm{u}}(T_{k,q}^{\mathrm{C}})$ can be expressed as $\bm{M}[\bm{x}_{\mathrm{u}}(T_{k,q}^{\mathrm{C}})]$ in (6). Note that, the CCM related to the estimation of channel $\bm{M}[\bm{x}_{\mathrm{u}}(T_{k,q}^{\mathrm{C}})]$ can be obtained by the probability density of $\bm{x}_{\mathrm{u}}(T_{k,q}^{\mathrm{C}})$ as well as the  mapping from the user location distribution to CCM, as shown from (8). However, obtaining the actual probability density of $\bm{x}_{\mathrm{u}}(T_{k,q}^{\mathrm{C}})$ is very difficult, because the user moving trajectory is hard to predict. For simplicity, we assume the user speed will not change during one COCT, i.e., $v_{\mathrm{u}}(T_{k}^{\mathrm{S}})=v_{\mathrm{u}}(T_{k}^{\mathrm{S}}+\tau)$ , $\forall\tau\in[0, T_{\mathrm{co}})$, and try to only utilize $\bm{x}_{\mathrm{u}}(T_{k}^{\mathrm{S}})$ and $v_{\mathrm{u}}(T_{k}^{\mathrm{S}})$ to approximately estimate the probability distribution of $\bm{x}_{\mathrm{u}}(T_{k,q}^{\mathrm{C}})$, $q=1,2,\cdots,N$.
\begin{figure}[t]
\centering
\includegraphics[width=0.7\textwidth]{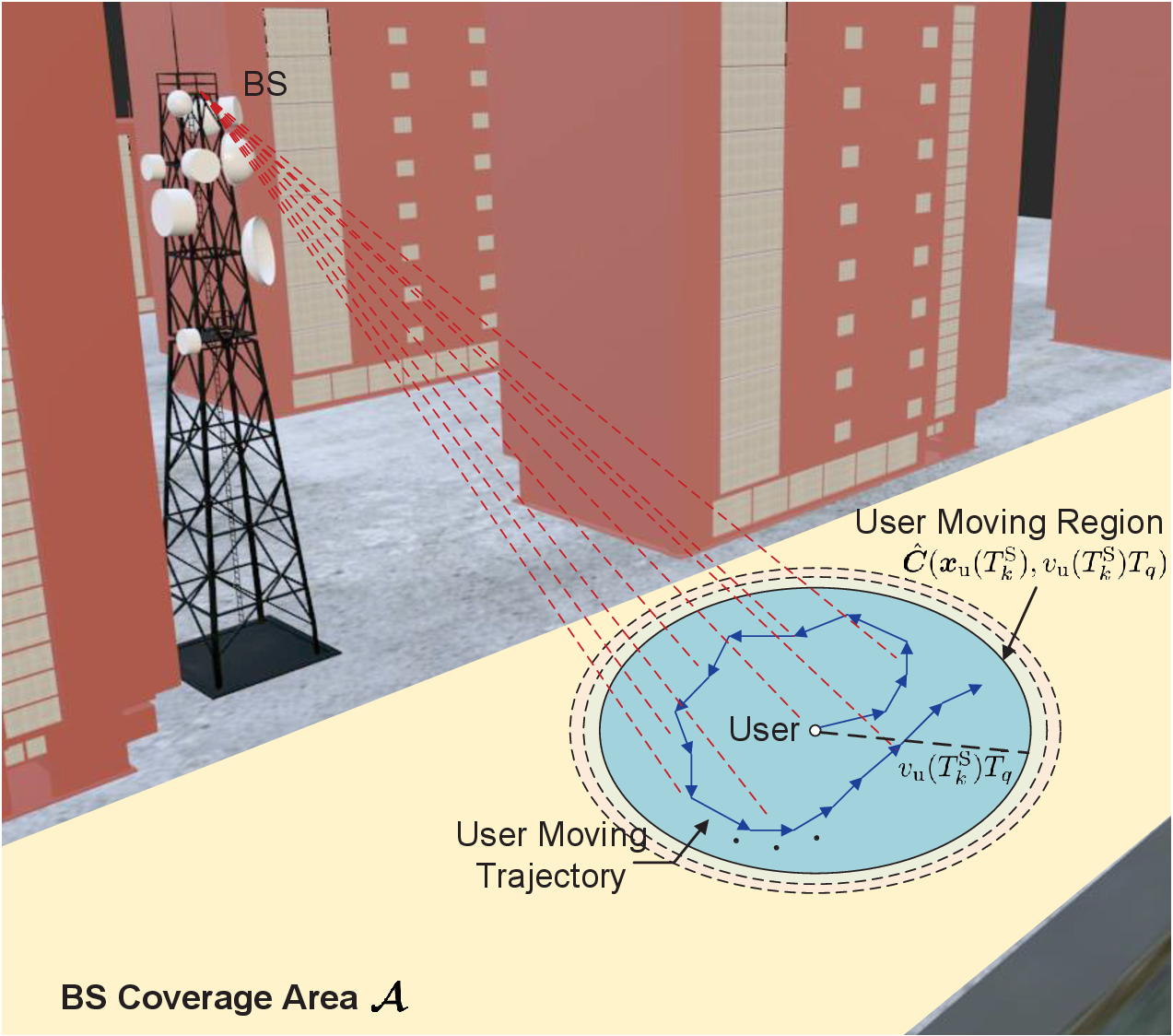}
\caption{The BS coverage area and user moving region.}
\end{figure}\\
\indent
As shown in Fig.~2, we consider the user only moves on a plane $\bm{\mathcal{A}}$. Since  $\bm{x}_{\mathrm{u}}(T_{k,q}^{\mathrm{C}})$, $k=1,2,\cdots$, $q=1,2,\cdots,N$, satisfies
\begin{equation}
||\bm{x}_{\mathrm{u}}(T_{k,q}^{\mathrm{C}})-\bm{x}_{\mathrm{u}}(T_{k}^{\mathrm{S}})||_2\leq v_{\mathrm{u}}(T_{k}^{\mathrm{S}})(T_{k,q}^{\mathrm{C}}-T_{k}^{\mathrm{S}}),
\end{equation}
the circle area $\hat{\bm{C}}(\bm{x}_{\mathrm{u}}(T_{k}^{\mathrm{S}}),v_{\mathrm{u}}(T_{k}^{\mathrm{S}})T_q)$, i.e., the user moving region in Fig.~2, would contain all possible values of $\bm{x}_{\mathrm{u}}(T_{k,q}^{\mathrm{C}})$, where $\hat{\bm{C}}(\hat{\bm{x}},\hat{r})=\{\bm{x}\ |\ ||\bm{x}-\hat{\bm{x}}||_2\leq \hat{r}, [\bm{x}]_3=H\}$, $H$ is the height coordinate of plane $\bm{\mathcal{A}}$, and $T_q=T_{k,q}^{\mathrm{C}}-T_{k}^{\mathrm{S}}$. Moreover, with no additional information, BS would reasonably assume the uniform distribution of $\bm{x}_{\mathrm{u}}(T_{k,q}^{\mathrm{C}})$ over $\hat{\bm{C}}(\bm{x}_{\mathrm{u}}(T_{k}^{\mathrm{S}}),v_{\mathrm{u}}(T_{k}^{\mathrm{S}})T_q)$. Then, the corresponding CCM for a given $\bm{x}_{\mathrm{u}}(T_{k,q}^{\mathrm{C}})$ can be represented as
\begin{equation}
\bm{R}_{k,q}=\bm{\psi}_q(\bm{x}_{\mathrm{u}}(T_{k}^{\mathrm{S}}),v_{\mathrm{u}}(T_{k}^{\mathrm{S}})),
\end{equation}
where
\begin{equation}
\begin{aligned}
&\bm{\psi}_q(\bm{x},v)=\\
&\frac{1}{\pi v^2T_q^2}\int_{0}^{2\pi}\int_{0}^{vT_q} \bm{M}[\bm{c}(\bm{x},\rho,\theta)]\bm{M}^{\mathrm{H}}[\bm{c}(\bm{x},\rho,\theta)]\rho\mathrm{d}\rho\mathrm{d}\theta
\end{aligned}
\end{equation}
and $\bm{c}(\bm{x},\rho,\theta)=[[\bm{x}]_1+\rho\cos(\theta),[\bm{x}]_2+\rho\sin(\theta),H]^{\mathrm{T}}$. By utilizing $\bm{R}_{k,q}$ to design pilot signals, the channel $\bm{M}[\bm{x}_{\mathrm{u}}(T_{k,q}^{\mathrm{C}})]$ can be estimated. However, the scale of the DNN for learning all the mapping $\bm{\psi}_q(\bm{x},v)$, $q=1,2,\cdots,N$, can be very large, which causes serious training and computation overhead. Therefore, we define the CCM $\bar{\bm{R}}_{k}$ to estimate all channels within one COCT as the average value
\begin{equation}
\begin{aligned}
\bar{\bm{R}}_{k}&=\bm{\psi}(\bm{x}_{\mathrm{u}}(T_{k}^{\mathrm{S}}),v_{\mathrm{u}}(T_{k}^{\mathrm{S}}))\\
&=\frac{1}{N}\sum_{q=1}^{N}\bm{\psi}_q(\bm{x}_{\mathrm{u}}(T_{k}^{\mathrm{S}}),v_{\mathrm{u}}(T_{k}^{\mathrm{S}}))=\frac{1}{N}\sum_{q=1}^{N}\bm{R}_{k,q}.
\end{aligned}
\end{equation}
Hence, once we obtained the mapping $\bm{\psi}(\bm{x},v)$, an effective CCM can be estimated in both the BS and user from $\bm{x}_{\mathrm{u}}(T_{k}^{\mathrm{S}})$ and $v_{\mathrm{u}}(T_{k}^{\mathrm{S}})$, as shown in (12).

\begin{figure}[t]
\centering
\includegraphics[width=0.75\textwidth]{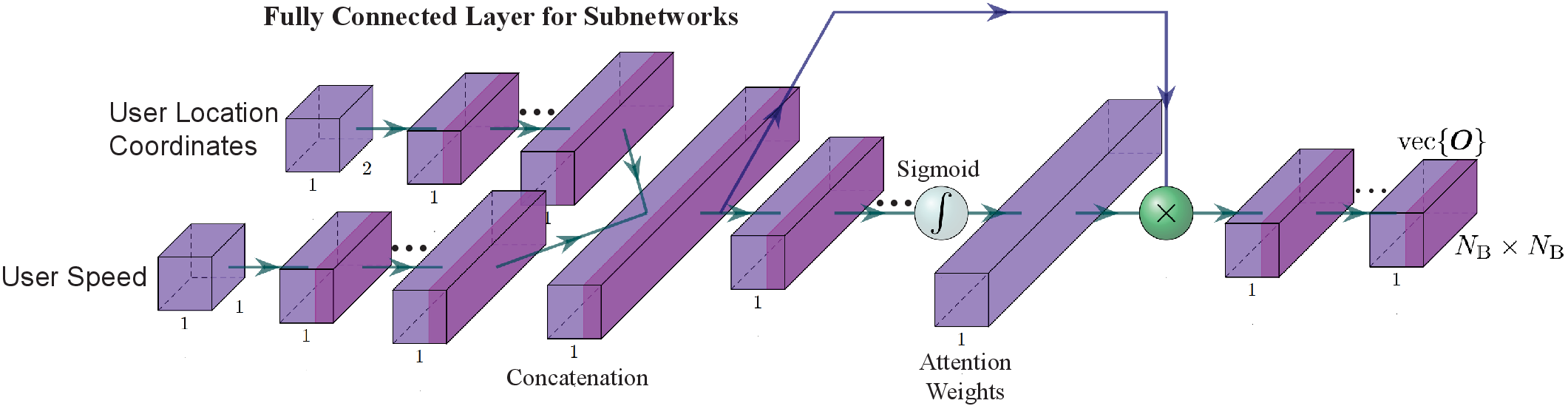}
\caption{The proposed architecture of LCNET.}
\end{figure}

Next we present the detailed design of DNN for the location based CCM estimation, called LCNET, where the feature fusion architecture \cite{mv3d} is adopted to approximately  learn the mapping $\bm{\psi}(\bm{x},v)$, as shown in Fig.~3. The proposed architecture contains different subnetworks to extract features from multiple inputs respectively and then concatenates all the extracted features as the overall input for the next network layer.
Moreover, we utilize the attention mechanism \cite{LChen}-\cite{Vaswani} that has the been validated as an effective technique to enhance the neural network performance. Specifically, we feed the concatenate layer into several fully connected layers and utilize the sigmoid function as the activation function of the last fully connected layer to obtain the attention weights. By the element-wise multiplication between the concatenate layer and the activation weights, the user location and speed can obtain attention with different levels to improve the learning accuracy.

Since $[\bm{x}]_3$ is a constant $H$, the input features of LCNET can be taken as the plane coordinates $[\bm{x}]_{1:2}\in\mathbb{R}^{2\times 1}$ of the user location $\bm{x}$ and user speed $v$. Then the output of LCNET is an $N_{\mathrm{B}}\times N_{\mathrm{B}}$ vector $\mathrm{vec}(\bm{O})$ with
\begin{equation}
[\bm{O}]_{i,j}=\left\{\begin{matrix}
               \mathrm{Re}\{[\bm{\psi}(\bm{x},v)]_{i,j}\} ,\ i\geq j \\
               \mathrm{Im}\{[\bm{\psi}(\bm{x},v)]_{i,j}\},\ i<j
              \end{matrix}\right..
\end{equation}

Since $\bm{\psi}(\bm{x},v)$ is conjugate symmetric, the output $\mathrm{vec}(\bm{O})$ of LCNET includes the real and imaginary part of all the elements of $\bm{\psi}(\bm{x},v)$ and thus can be utilized to recover $\bm{\psi}(\bm{x},v)$. It is worth noting that though the dimensions of $\bm{\psi}(\bm{x},v)$ are generally much larger than the dimensions of $[\bm{x}]_{1:2}$, we still use one DNN to predict the real part and imaginary part of $\bm{\psi}(\bm{x},v)$ to learn the correlation between different elements of $\bm{\psi}(\bm{x},v)$. Moreover, as the estimated CCM from $\mathrm{vec}(\bm{O})$ may not guarantee its positive semi-definiteness, all the negative eigenvalues of the estimated CCM will be replaced by the minimum nonnegative eigenvalue.

\subsection{Enhanced CCM Estimation with Denoising}
In practice, the reported user locations would inevitably contain errors, which will degrade the performance of ULCCE. We propose that BS computes a historical user location from the previously estimated channel and then performs denoising for the current location by the Kalman framework. For the $k$th COCT, we know $\bm{x}_{\mathrm{u}}(T_{k,N}^{\mathrm{C}})$ is the user location of the $N$th CCT and can be very close to the user location $\bm{x}_{\mathrm{u}}(T_{k+1}^{\mathrm{S}})$ that will be fed back in the next COCT. Thus, BS can utilize the inverse mapping $\bm{M}^{-1}$ to estimate $\bm{x}_{\mathrm{u}}(T_{k,N}^{\mathrm{C}})$ and then  improve the estimation accuracy of $\bm{x}_{\mathrm{u}}(T_{k+1}^{\mathrm{S}})$.

Specifically, we design a location estimation DNN, called LENET, with multiple fully connected layers to learn the mapping $\bm{M}^{-1}$, as shown in Fig.~4. The input feature of LENET is chosen as the vector concatenated by the normalized real part and the normalized image part of one channel $\bm{h}$, i.e., $[\frac{1}{\zeta}\mathrm{Re}(\bm{h})^{\mathrm{T}},\frac{1}{\zeta}\mathrm{Im}(\bm{h})^{\mathrm{T}}]^{\mathrm{T}}$, where $\zeta=\mathrm{E}\{||\bm{h}||_2\}$ is the average channel amplitude. The output of LENET is the plane coordinates $[\bm{x}]_{1:2}$ of the user location corresponding to the channel $\bm{h}$. It is worth noting that the channel of each sample in the training set of LENET is considered to be corrupted by the Gaussian noise $\tilde{\bm{n}}\in \mathcal{CN}(\bm{0},\tilde{\sigma}^2\bm{I}_{N_{\mathrm{B}}})$ in order to enhance the robustness of LENET. Once LENET is well trained, BS can estimate $\bm{x}_{\mathrm{u}}(T_{k,N}^{\mathrm{C}})$ at the $k$th COCT for any $k$.
\begin{figure}[t]
\centering
\includegraphics[width=0.75\textwidth]{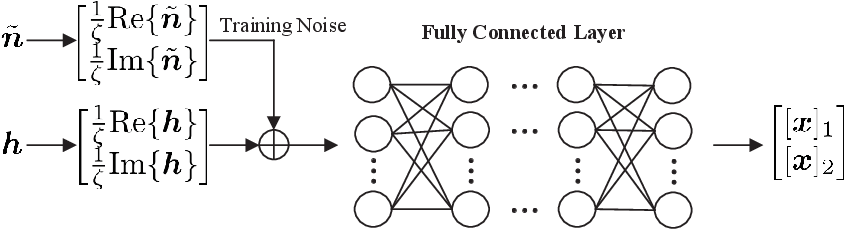}
\caption{The proposed architecture of LENET.}
\end{figure}

Denote the estimated channel of the $N$th CCT at the $k$th COCT as $\bm{h}_{k,N}$. Since $\bm{h}_{k,N}$ is a random vector, the location estimation error of LENET $\bm{n}_k=\hat{\bm{M}}^{-1}(\bm{h}_{k,N})-\tilde{\bm{x}}_{\mathrm{u}}(T_{k,N}^{\mathrm{C}})$ will be a two dimensional real random vector, where $\hat{\bm{M}}^{-1}(\cdot)$ denotes the learned mapping of LENET and $\tilde{\bm{x}}_{\mathrm{u}}(t)=[[\bm{x}_{\mathrm{u}}(t)]_1,[\bm{x}_{\mathrm{u}}(t)]_2]^{\mathrm{T}}$ is the plane coordinates of $\bm{x}_{\mathrm{u}}(t)$. As the probability distribution of $\bm{n}_k$ is hard to determine, we use Gaussian distribution $\mathcal{N}(\bm{m}_{\mathrm{n}},\bm{R}_{\mathrm{n}})$ to approximate the actual probability distribution of $\bm{n}_k$, and then estimate $\bm{m}_{\mathrm{n}}$ and $\bm{R}_{\mathrm{n}}$ by an additional sample set of LENET. Specifically, by utilizing the additional sample set $\{(\bm{h}_{\mathrm{V},i},\bm{x}_{\mathrm{V},i})\ |\ i=1,2,\cdots,W\}$, $\bm{m}_{\mathrm{n}}$ and $\bm{R}_{\mathrm{n}}$ can be derived as
\begin{equation}
\begin{aligned}
\bm{m}_{\mathrm{n}}&=\frac{1}{WQ}\sum_{i=1}^{W}\sum_{j=1}^{Q}\bm{n}_{\mathrm{V},i,j},\\
\bm{R}_{\mathrm{n}}&=\frac{1}{WQ-1}\sum_{i=1}^{W}\sum_{j=1}^{Q}(\bm{n}_{\mathrm{V},i,j}-\bm{m}_{\mathrm{n}})(\bm{n}_{\mathrm{V},i,j}-\bm{m}_{\mathrm{n}})^{\mathrm{T}},
\end{aligned}
\end{equation}
where
\begin{equation}
\bm{n}_{\mathrm{V},i,j}=\hat{\bm{M}}^{-1}(\bm{h}_{\mathrm{V},i}+\tilde{\bm{n}}_{i,j})-\tilde{\bm{x}}_{\mathrm{V},i},
\end{equation}
$\bm{h}_{\mathrm{V},i}$ is the accurate sample channel as input feature, $\tilde{\bm{x}}_{\mathrm{V},i}\in \mathbb{R}^{2\times1}$ is the corresponding user plane coordinates as sample label, and $\tilde{\bm{n}}_{i,j}$, $i=1,2,\cdots,W$, $j=1,2,\cdots,Q$, is a sample from $\mathcal{CN}(\bm{0},\tilde{\sigma}^2\bm{I}_{N_{\mathrm{B}}})$.
\begin{figure*}[t]
\centering
\includegraphics[width=1\textwidth]{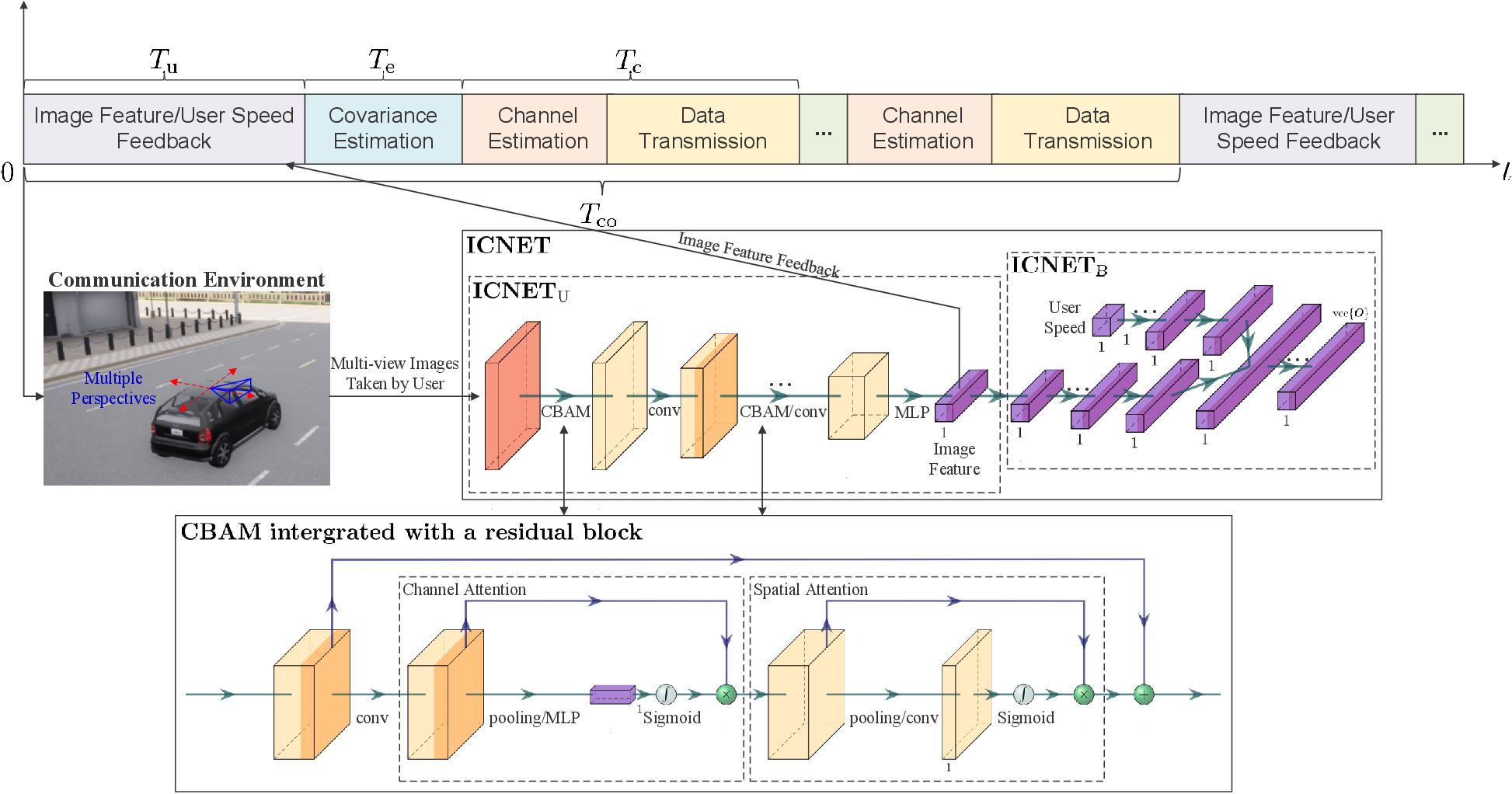}
\caption{The timing diagram of the proposed image based covariance estimation.}
\end{figure*}

As $\hat{\bm{M}}^{-1}(\bm{h}_{k,N})$ is the estimate of $\tilde{\bm{x}}_{\mathrm{u}}(T_{k,N}^{\mathrm{C}})$, the distribution of $\tilde{\bm{x}}_{\mathrm{u}}(T_{k+1}^{\mathrm{S}})-\tilde{\bm{x}}_{\mathrm{u}}(T_{k,N}^{\mathrm{C}})$ needs to be obtained to estimate $\tilde{\bm{x}}_{\mathrm{u}}(T_{k+1}^{\mathrm{S}})$ from $\hat{\bm{M}}^{-1}(\bm{h}_{k,N})$. Since $||\tilde{\bm{x}}_{\mathrm{u}}(T_{k+1}^{\mathrm{S}})-\tilde{\bm{x}}_{\mathrm{u}}(T_{k,N}^{\mathrm{C}})||_2\leq v_{\mathrm{u}}(T_{k}^{\mathrm{S}})T_{\mathrm{c}}$ and the general user speed can only lead to a very small displacement during milliseconds, i.e., $v_{\mathrm{u}}(T_{k}^{\mathrm{S}})T_{\mathrm{c}}\ll 1$, we may assume $\tilde{\bm{x}}_{\mathrm{u}}(T_{k+1}^{\mathrm{S}})-\tilde{\bm{x}}_{\mathrm{u}}(T_{k,N}^{\mathrm{C}})\approx \bm{0}$. Thus, $\hat{\bm{M}}^{-1}(\bm{h}_{k,N})$ can be approximated as an estimate of $\tilde{\bm{x}}_{\mathrm{u}}(T_{k+1}^{\mathrm{S}})$ and obeys $\mathcal{N}(\tilde{\bm{x}}_{\mathrm{u}}(T_{k+1}^{\mathrm{S}})+\bm{m}_{\mathrm{n}},\bm{R}_{\mathrm{n}})$. We further assume the estimation noise of $\tilde{\bm{x}}_{\mathrm{u}}(T_{k+1}^{\mathrm{S}})$ will follow the Gaussian distribution $\mathcal{N}(\bm{0},\sigma_{\mathrm{c}}^2\bm{I}_2)$ and denote the estimate of $\tilde{\bm{x}}_{\mathrm{u}}(T_{k+1}^{\mathrm{S}})$ as $\breve{\bm{x}}_{k+1}\in \mathbb{R}^{2\times1}$, namely the distribution of $\breve{\bm{x}}_{k+1}$ is $\mathcal{N}(\tilde{\bm{x}}_{\mathrm{u}}(T_{k+1}^{\mathrm{S}}),\sigma_{\mathrm{c}}^2\bm{I}_2)$.

Note that $\hat{\bm{M}}^{-1}(\bm{h}_{k,N})$ and $\breve{\bm{x}}_{k+1}$ are both the measurements of $\tilde{\bm{x}}_{\mathrm{u}}(T_{k+1}^{\mathrm{S}})$ under the Kalman framework. Moreover, the distribution of $\tilde{ \bm{x}}_{\mathrm{u}}(T_{k+1}^{\mathrm{S}})$ can by predicted by the distribution of $\tilde{\bm{x}}_{\mathrm{u}}(T_{k}^{\mathrm{S}})$ and the uploaded speed $v_{\mathrm{u}}(T_k^{\mathrm{S}})$. Specifically, as $\|\tilde{\bm{x}}_{\mathrm{u}}(T_{k+1}^{\mathrm{S}})-\tilde{\bm{x}}_{\mathrm{u}}(T_{k}^{\mathrm{S}})\|_{2}\leq v_{\mathrm{u}}(T_k^{\mathrm{S}})T_{\mathrm{co}}$, we can assume $\tilde{\bm{u}}_k=\tilde{\bm{x}}_{\mathrm{u}}(T_{k+1}^{\mathrm{S}})-\tilde{\bm{x}}_{\mathrm{u}}(T_{k}^{\mathrm{S}})$ will obey two dimensional uniform distribution over the circle area $\{\bm{u}\ |\ \|\bm{u}\|_2\leq v_{\mathrm{u}}(T_k^{\mathrm{S}})T_{\mathrm{co}}\}$ and is independent from $\tilde{\bm{x}}_{\mathrm{u}}(T_{k}^{\mathrm{S}})$. Then, the prediction step for $\tilde{\bm{x}}_{\mathrm{u}}(T_{k+1}^{\mathrm{S}})$ can be expressed as
\begin{equation}
\tilde{\bm{x}}_{\mathrm{u}}(T_{k+1}^{\mathrm{S}})=\tilde{\bm{x}}_{\mathrm{u}}(T_{k}^{\mathrm{S}})+\tilde{\bm{u}}_{k}.
\end{equation}
Without loss of generality, $\tilde{\bm{x}}_{\mathrm{u}}(T_{k}^{\mathrm{S}})$ can be assumed to obey $\mathcal{N}(\bm{m}^{\mathrm{S}}_k,\bm{R}^{\mathrm{S}}_k)$. From (16), the probability distribution function of $\tilde{\bm{x}}_{\mathrm{u}}(T_{k+1}^{\mathrm{S}})$ will not be Gaussian distribution but the convolution of a Gaussian distribution and a uniform distribution. The non-Gaussian distribution of $\tilde{\bm{x}}_{\mathrm{u}}(T_{k+1}^{\mathrm{S}})$ can not meet the requirements of iteration of Kalman filter. Here, we approximately replace $\tilde{\bm{u}}_{k}$ in (16) with a Gaussian variable $\bm{n}^{\mathrm{U}}_{k}$, where $\bm{n}^{\mathrm{U}}_{k}$ has the same mean value and covariance of $\tilde{\bm{u}}_{k}$, i.e., $\bm{n}^{\mathrm{U}}_{k}\in \mathcal{N}(\bm{0},\bm{R}^{\mathrm{U}}_k)$ and $\bm{R}^{\mathrm{U}}_k=\frac{v^2_{\mathrm{u}}(T_k^{\mathrm{S}})T^2_{\mathrm{co}}}{4}\bm{I}_2$. Thus, the prior probability distribution of $\tilde{\bm{x}}_{\mathrm{u}}(T_{k+1}^{\mathrm{S}})$ can be given by $\mathcal{N}(\bm{m}^{\mathrm{S}}_k,\bm{R}^{\mathrm{S}}_k+\bm{R}^{\mathrm{U}}_k)$.

By assuming $\hat{\bm{M}}^{-1}(\bm{h}_{k,N})$ and $\breve{\bm{x}}_{k}$ are conditionally independent for a certain $\tilde{\bm{x}}_{\mathrm{u}}(T_{k+1}^{\mathrm{S}})$, the conditional PDF $p(\tilde{\bm{x}}_{\mathrm{u}}(T_{k+1}^{\mathrm{S}})\ |\ \hat{\bm{M}}^{-1}(\bm{h}_{k,N}),\breve{\bm{x}}_{k+1})$ is given by
\begin{equation}
\begin{aligned}
&p(\tilde{\bm{x}}_{\mathrm{u}}(T_{k+1}^{\mathrm{S}})\ |\ \hat{\bm{M}}^{-1}(\bm{h}_{k,N}),\breve{\bm{x}}_{k})\\
&=\frac{p(\hat{\bm{M}}^{-1}(\bm{h}_{k,N})\ |\ \tilde{\bm{x}}_{\mathrm{u}}(T_{k+1}^{\mathrm{S}}))p(\breve{\bm{x}}_{k}\ |\ \tilde{\bm{x}}_{\mathrm{u}}(T_{k+1}^{\mathrm{S}}))p(\tilde{\bm{x}}_{\mathrm{u}}(T_{k+1}^{\mathrm{S}}))}{p(\hat{\bm{M}}^{-1}(\bm{h}_{k,N}),\breve{\bm{x}}_{k})}\\
&
\begin{aligned}
=&\frac{|\bm{R}_{\mathrm{n}}(\bm{R}^{\mathrm{S}}_k+\bm{R}_k^{\mathrm{U}})|^{-\frac{1}{2}}}{8\pi^3\sigma_{\mathrm{c}}^2p(\hat{\bm{M}}^{-1}(\bm{h}_{k,N}),\breve{\bm{x}}_{k})}e^{-\frac{[\tilde{\bm{x}}_{\mathrm{u}}(T_{k+1}^{\mathrm{S}})-\breve{\bm{x}}_{k+1}]^{\mathrm{T}}[\tilde{\bm{x}}_{\mathrm{u}}(T_{k+1}^{\mathrm{S}})-\breve{\bm{x}}_{k+1}]}{2\sigma_c^2}}\\
&\cdot e^{-\frac{1}{2}(\tilde{\bm{x}}_{\mathrm{u}}(T_{k+1}^{\mathrm{S}})-\bm{m}^{\mathrm{S}}_k)^{\mathrm{T}}(\bm{R}^{\mathrm{S}}_k+\bm{R}_k^{\mathrm{U}})^{-1}(\tilde{\bm{x}}_{\mathrm{u}}(T_{k+1}^{\mathrm{S}})-\bm{m}^{\mathrm{S}}_k)}\\
&\cdot
e^{-\frac{1}{2}(\tilde{\bm{x}}_{\mathrm{u}}(T_{k+1}^{\mathrm{S}})-{\bm{x}}_{k}^{\mathrm{C}})^{\mathrm{T}}\bm{R}_{\mathrm{n}}^{-1}(\tilde{\bm{x}}_{\mathrm{u}}(T_{k+1}^{\mathrm{S}})-{\bm{x}}_{k}^{\mathrm{C}})}
\end{aligned}\\
&=\frac{1}{2\pi\sqrt{|\bm{R}^{\mathrm{KM}}_{k+1}|}}e^{-\frac{1}{2}(\tilde{\bm{x}}_{\mathrm{u}}(T_{k+1}^{\mathrm{S}})-\bm{m}^{\mathrm{KM}}_{\mathrm{k+1}})^{\mathrm{T}}(\bm{R}^{\mathrm{KM}}_{\mathrm{k+1}})^{-1}(\tilde{\bm{x}}_{\mathrm{u}}(T_{k+1}^{\mathrm{S}})-\bm{m}^{\mathrm{KM}}_{\mathrm{k+1}})},
\end{aligned}
\end{equation}
where,
\begin{equation}
\begin{aligned}
{\bm{x}}_{k}^{\mathrm{C}}&=\hat{\bm{M}}^{-1}(\bm{h}_{k,N})-\bm{m}_{\mathrm{n}},\\
\bm{m}^{\mathrm{KM}}_{k+1}&=\bm{R}^{\mathrm{KM}}_{k+1}(\bm{R}_{\mathrm{n}}^{-1}\tilde{\bm{x}}_{k}^{\mathrm{C}}+\sigma_{\mathrm{c}}^{-2}\breve{\bm{x}}_{k+1}+(\bm{R}^{\mathrm{S}}_k+\bm{R}_k^{\mathrm{U}})^{-1}\bm{m}^{\mathrm{S}}_k),\\
\bm{R}^{\mathrm{KM}}_{k+1}&=(\sigma_{\mathrm{c}}^{-2}\bm{I}_2+\bm{R}_{\mathrm{n}}^{-1}+(\bm{R}^{\mathrm{S}}_k+\bm{R}_k^{\mathrm{U}})^{-1})^{-1},
\end{aligned}
\end{equation}
$p(\hat{\bm{M}}^{-1}(\bm{h}_{k,N})\ |\ \tilde{\bm{x}}_{\mathrm{u}}(T_{k+1}^{\mathrm{S}}))$, $p(\breve{\bm{x}}_{k}\ |\ \tilde{\bm{x}}_{\mathrm{u}}(T_{k+1}^{\mathrm{S}}))$ and $p(\tilde{\bm{x}}_{\mathrm{u}}(T_{k+1}^{\mathrm{S}}))$ are the PDF of $\mathcal{N}(\tilde{\bm{x}}_{\mathrm{u}}(T_{k+1}^{\mathrm{S}})+\bm{m}_{\mathrm{n}},\bm{R}_{\mathrm{n}})$, $\mathcal{N}(\tilde{\bm{x}}_{\mathrm{u}}(T_{k+1}^{\mathrm{S}}),\sigma_{\mathrm{c}}^2\bm{I}_2)$ and $\mathcal{N}(\bm{m}^{\mathrm{S}}_k,\bm{R}^{\mathrm{S}}_k+\bm{R}^{\mathrm{U}}_k)$ respectively.

The conditional mean value $\bm{m}^{\mathrm{KM}}_{\mathrm{k+1}}$ can be treated as the corrected estimation of $\tilde{\bm{x}}_{\mathrm{u}}(T_{k+1}^{\mathrm{S}})$. For the $(k+1)$th COCT, $\bm{m}^{\mathrm{KM}}_{\mathrm{k+1}}$ and $\bm{R}^{\mathrm{KM}}_{\mathrm{k+1}}$ can be further used as the new mean and covariance of the prior probability distribution of $\tilde{\bm{x}}_{\mathrm{u}}(T_{k+1}^{\mathrm{S}})$, i.e., $\bm{m}^{\mathrm{S}}_{k+1}=\bm{m}^{\mathrm{KM}}_{k+1}$ and $\bm{R}^{\mathrm{S}}_{k+1}=\bm{R}^{\mathrm{KM}}_{k+1}$, to implement iterations of Kalman filter. Then, the enhanced CCM estimation method with denoising is shown in Algorithm 1.
\begin{algorithm}
\caption{The enhanced CCM estimation method with denoising}
\label{alg1}
\begin{algorithmic}[1]
\REQUIRE ~~\\
The variance of positioning noise, $\sigma_{\mathrm{c}}^2$;\\
\ENSURE ~~\\
The sequence of the estimated covariance matrices of all COCTs;
\STATE $k=1$;
\STATE User feeds back $\breve{\bm{x}}_{k}$ and $v_{\mathrm{u}}(T_k^{\mathrm{S}})$ to the BS;
\STATE BS and user both utilize LCNET to estimate the covariance $\bm{\psi}(\breve{\bm{x}}_{k},v_{\mathrm{u}}(T_k^{\mathrm{S}}))$;
\STATE BS utilizes the estimated covariance from LCNET to design pilot signals to estimate channel $\bm{M}[\bm{x}_{\mathrm{u}}(T_{k,q}^{\mathrm{C}})]$, $q=1,2,\cdots,N$, at user end;
\STATE User feeds back the estimated channel $\bm{h}_{k,N}$ or feeds back $\hat{\bm{M}}^{-1}(\bm{h}_{k,N})$ to BS based on the saved LENET;
\STATE Initialize $\bm{m}^{\mathrm{S}}_k=\breve{\bm{x}}_{k}$ and $\bm{R}^{\mathrm{S}}_k=\sigma_c^2\bm{I}_2$ for location denoising;
\REPEAT
\STATE User feeds back $\breve{\bm{x}}_{k+1}$ and $v_{\mathrm{u}}(T_{k+1}^{\mathrm{S}})$ to the BS;
\STATE Obtain $\bm{m}^{\mathrm{KM}}_{k+1}$ and $\bm{R}^{\mathrm{KM}}_{k+1}$ from (18) by using $\breve{\bm{x}}_{k+1}$, $v_{\mathrm{u}}(T_k^{\mathrm{S}})$, $\hat{\bm{M}}^{-1}(\bm{h}_{k,N})$, $\bm{m}^{\mathrm{S}}_k$ and $\bm{R}^{\mathrm{S}}_k$;
\STATE BS and user both utilize LCNET to estimate the covariance $\bm{\psi}(\bm{m}^{\mathrm{KM}}_{k+1},v_{\mathrm{u}}(T_{k+1}^{\mathrm{S}}))$ for channel estimation;
\STATE User feeds back the estimated channel $\bm{h}_{k+1,N}$ or $\hat{\bm{M}}^{-1}(\bm{h}_{k+1,N})$ to BS;
\STATE $\bm{m}^{\mathrm{S}}_{k+1}=\bm{m}^{\mathrm{KM}}_{k+1}$ and $\bm{R}^{\mathrm{S}}_{k+1}=\bm{R}^{\mathrm{KM}}_{k+1}$;
\STATE $k:=k+1$;
\UNTIL{End of communication}
\end{algorithmic}
\end{algorithm}

\begin{figure}[t]
\centering
\includegraphics[width=0.7\textwidth]{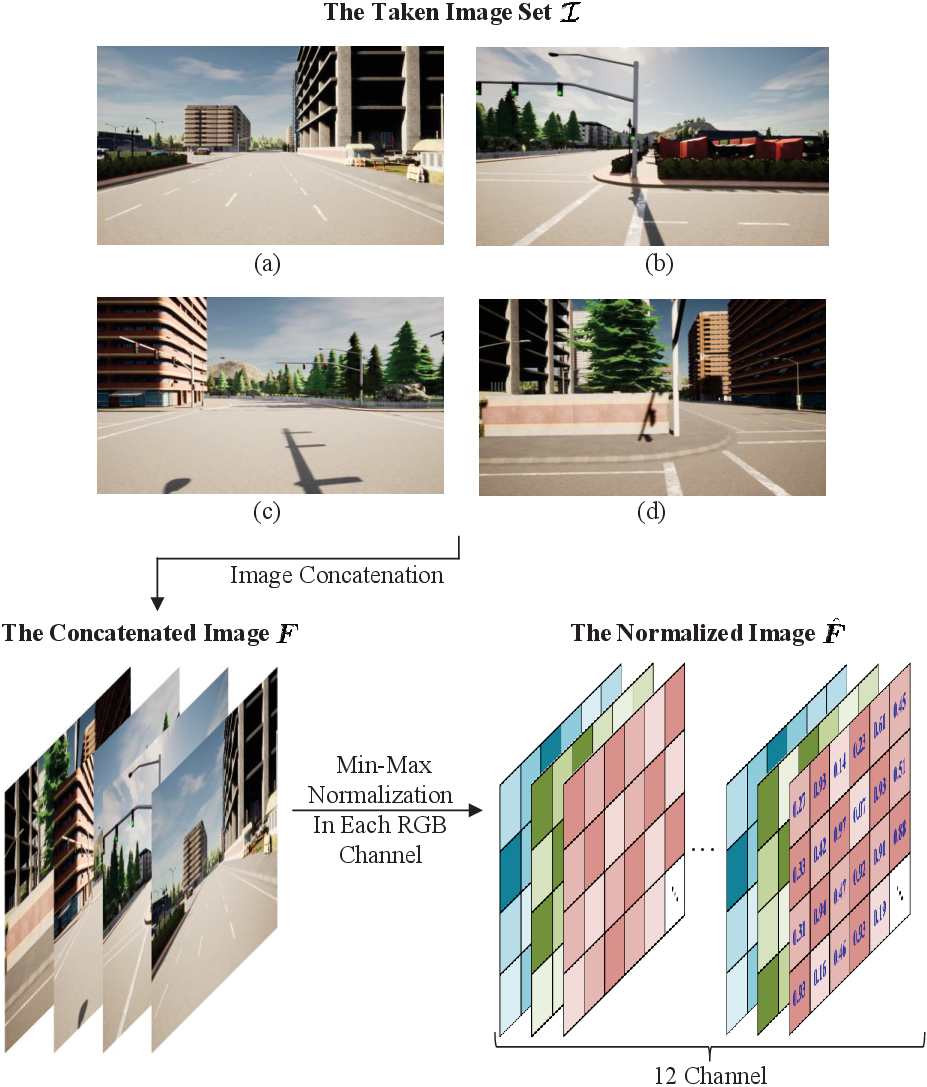}
\caption{The taken image set $\bm{\mathcal{I}}$ and the input feature generation of $\mathrm{ICNET}_{\mathrm{U}}$ from $\bm{\mathcal{I}}$.}
\end{figure}

\section{Scene Image Based CCM Estimation Method}
Practically, the user location will be unavailable due to the strong noise/interference corruption or the user's privacy protection. In these cases, we propose to use scene image taken from user terminal (not BS) to estimate the CCM. Note that, camera has been widely used as an auxiliary equipment for many mobile intelligent terminals, such as autonomous vehicles, unmanned aerial vehicles or even the smart phone. It is expected that scene images taken by user can reflect the user location information and represent spatial characteristics  between the BS and user. Fig.~5 shows the frame structure of the proposed SICCME method. Compared with ULCCME method, the user will feed back its speed and the image feature rather than its location. Then, BS will utilize these information to estimate CCM during each COCT.

Specifically, we design an image based DNN for CCM estimation, called ICNET, with the feature fusion architecture to learn the mapping $\bm{\psi}_{\mathrm{I}}(\bm{\mathcal{I}}, v):\ (\bm{\mathcal{I}},v)\rightarrow \bm{\psi}(\bm{x},v)$ as shown in Fig.~5, where $\bm{\mathcal{I}}$ is the set of scene images taken by user at location $\bm{x}$ from multiple camera angles. We divide the ICNET into two parts: ICNET$_\textrm{U}$ and ICNET$_\textrm{B}$. ICNET$_\textrm{U}$ is the partial subnetwork corresponding to the input images $\bm{\mathcal{I}}$ and is used for image feature extraction. As shown in Fig.~6, different images in $\bm{\mathcal{I}}$ can contain different spatial information and there may be some images that contain much richer and more useful spatial information than others. For instance, the Fig.~6(b) and Fig.~6(d) correspond the sides of the road and can contain more scene objects or buildings than the Fig.~6(a) and Fig.~6(c). Hence, we utilize the convolutional block attention module (CBAM) \cite{Sanghyun} integrated with residual network architecture \cite{HeKai} to design ICNET$_\textrm{U}$. The CBAM includes spatial attention and channel attention mechanism and thus can provide the different parts or color channels of the images in $\bm{\mathcal{I}}$ with different degrees of attention. The output of ICNET$_\textrm{U}$, i.e., the bottleneck between ICNET$_\textrm{U}$ and ICNET$_\textrm{B}$, is the extracted image feature. Since the dimension of image feature vector is much lower than the original images, ICNET$_\textrm{U}$ can be saved at user terminal to avoid the need of directly feeding back the image. For a fair comparison with ULCCME, the dimension of image feature is set as 2 here to make the feedback overhead of SICCME consistent with ULCCME. ICNET$_\textrm{B}$ is used to estimate CCM from the uploaded image feature and user speed and stored in BS. During the practical implementation, ICNET$_\textrm{U}$ and ICNET$_\textrm{B}$ should be jointly trained. Once ICNET is well
trained, ICNET$_\textrm{U}$ and ICNET$_\textrm{B}$ with well-trained weights are stored in user side and BS, respectively.

\begin{figure}[t]
\centering
\includegraphics[width=0.55\textwidth]{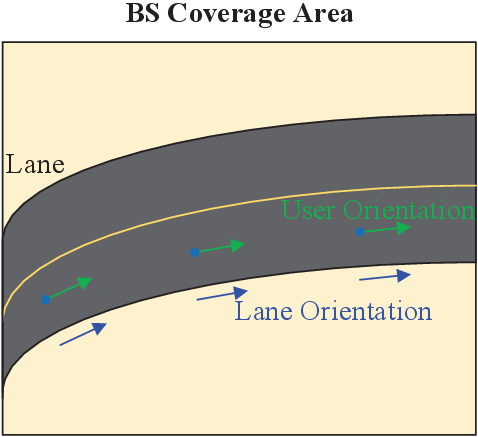}
\caption{The user orientation for the traffic lane.}
\end{figure}

In summary, before BS starts to estimate CCM $\bar{\bm{R}}_k$ for the $k$th COCT, the user will be asked to take images at the moment $T_{k}^{\mathrm{S}}$ and only needs to feed back the image feature obtained by ICNET$_\textrm{U}$. Then, the CCM $\bar{\bm{R}}_k$ can be estimated through the ICNET$_\textrm{B}$.

Since more images may more accurately reflect the user location and the surrounding environment, the performance of SICCME would depend on the number and camera angles of the scene images in $\bm{\mathcal{I}}$. We here illustrate an example when the user takes four pictures along the positive X-axis, negative X-axis, positive Y-axis and negative Y-axis of plane $\bm{\mathcal{A}}$ respectively. As shown in Fig.~6, we concatenate the four images in the dimension of RGB channel to obtain a 12-channel image and then normalize each channel of the 12-channel image to generate the input feature of ICNET$_\textrm{U}$. Specifically, the set $\bm{\mathcal{I}}$ will contain four images $\bm{F}_i\in\mathbb{R}^{f_{\mathrm{H}}\times f_{\mathrm{W}}\times 3}$, $i=1,2,3,4$ with RGB channels. We then obtain a 12-channel image $\bm{F}$ from $[\bm{F}]_{:,:,(3i-2):3i}=\bm{F}_i$, $i=1,2,3,4$. To enhance the learning performance of ICNET, we utilize the min-max way to normalize the $j$th channel of $\bm{F}$ as
\begin{equation}
\hat{\bm{F}}_{j}=\frac{[\bm{F}]_{:,:,j}-F_{\mathrm{min,j}}}{F_{\mathrm{max,j}}-F_{\mathrm{min,j}}},\ j=1,2,\cdots,12,
\end{equation}
where $F_{\mathrm{max,j}}/F_{\mathrm{min,j}}$ is the maximum/minimize pixel value in the $j$th channel of the concatenate image $\bm{F}$ of all ICNET's training samples. Then, we utilize the normalized image $\hat{\bm{F}}$ as the input feature, where $[\hat{\bm{F}}]_{:,:,j}=\hat{\bm{F}}_{j}$, $j=1,2,\cdots,12$. The design of output of ICNET is the same with LCNET as shown in (13), and the strategy to guarantee the positive semi-definiteness of the covariance acquired by ICNET is the same with ULCCME.

It is worth noting that though we here set the user to take images with four fixed camera shooting angles, i.e., the positive X-axis, negative X-axis, positive Y-axis and negative Y-axis of plane $\bm{\mathcal{A}}$, under different locations, the camera angles can be different for different locations in practice, as long as the shooting angle of each camera is unique for each location to ensure that there is only an image set corresponding to a location. For instance, when the user, i.e, a vehicle, runs along a lane in Fig.~7, the user orientation as well as the angles of the cameras of the user will change with the lane orientation, but the proposed SICCME is still applicable, since for each location at the lane, the vehicle orientation is usually parallel to the lane orientation. However, there must exist the exceptions when the vehicle does not run along the traffic lane. For these exceptions, we consider many techniques of autonomous driving, such as the lane detection \cite{Neven}, can be used to determine whether the vehicle orientation is consistent with the lane orientation, and thereby help to determine when to use the proposed SICCME.

\section{Simulation Results}
\subsection{Performance Metric}
The normalized mean-square error (NMSE) of CCM estimation is used to evaluate the learning accuracy of ULCCME and SICCME, defined as
\begin{equation}
\mathrm{NMSE}_{\mathrm{R}}=\frac{\sum_{i=1}^{S_{\mathrm{R}}}||\bm{R}_{i}-\hat{\bm{R}}_{i}||_F^2}{\sum_{i=1}^{S_{\mathrm{R}}}||\bm{R}_{i}||_F^2},
\end{equation}
where $\hat{\bm{R}}_k$ is the estimate of $\bm{R}_k$ through LCNET or ICNET, and $S_{\mathrm{R}}$ is the number of test samples of LCNET or ICNET.\\
\indent
After obtaining $\hat{\bm{R}}_{k}$, the linear minimum mean-square error (LMMSE) channel estimation \cite{Biguesh} is adopted for channel estimation within any CCT. By defining the eigen-decompostion of $\hat{\bm{R}}_{k}$ as $\hat{\bm{R}_k}=\bm{\Gamma}\bm{\Lambda}\bm{\Gamma}^{\mathrm{H}}$, we can design the pilot signal matrix for one COCT as
\begin{equation}
\bm{P}_{\mathrm{pilot}}(\hat{\bm{R}}_k)=\sigma\bm{\Gamma}[([\mu\bm{I}-\bm{\Lambda}^{-1}]^+)^{\frac{1}{2}},\bm{0}_{N_{\mathrm{B}}\times (M_{\mathrm{p}}-N_{\mathrm{B}})}]\bm{U},
\end{equation}
where $\bm{U}\in \mathbb{C}^{M_{\mathrm{p}}\times M_{\mathrm{p}}}$ can be any unitary matrix and $M_{\mathrm{p}}$ is the affordable length of pilot signals. Moreover, $\mu$ can be  derived from the well-known water-filling approach \cite{Gao} to satisfy the transmit energy constraint  $\mathrm{tr}(\bm{P}_{\mathrm{pilot}}(\hat{\bm{R}}_k)\bm{P}_{\mathrm{pilot}}^{\mathrm{H}}(\hat{\bm{R}}_k))=P$. Then, the LMMSE channel estimator at the receiver for the $k$th COCT is expressed as
\begin{equation}
\bm{A}_{\mathrm{opt}}(\hat{\bm{R}}_k)=(\bm{P}_{\mathrm{pilot}}^{\mathrm{H}}\hat{\bm{R}}_k\bm{P}_{\mathrm{pilot}}+\sigma^2\bm{I})^{-1}\bm{P}_{\mathrm{pilot}}^{\mathrm{H}}\hat{\bm{R}}_k.
\end{equation}
The NMSE of channel estimation is similarly defined as
\begin{equation}
\mathrm{NMSE}_{\mathrm{H}}=\frac{\sum_{i=1}^{S_{\mathrm{H}}}\mathrm{E}\{||\tilde{\bm{h}}_{i}-\hat{\bm{h}}_{i}||^2_2\}}{\sum_{i=1}^{S_{\mathrm{H}}}||\tilde{\bm{h}}_{i}||^2_2},
\end{equation}
where $\hat{\bm{h}}_i$ is the estimate of $\tilde{\bm{h}}_i$, and $S_{\mathrm{H}}$ is the number of tested channels to evaluate the channel estimation performance.

Moreover, the location estimation accuracy of LENET is evaluated by the root mean-square error (RMSE) performance, which is expressed as
\begin{equation}
\mathrm{RMSE}_{\mathrm{L}}=\sqrt{\frac{1}{2WQ}\sum_{i=1}^{W}\sum_{j=1}^{Q}||\bm{n}_{\mathrm{V},i,j}||_2^2}.
\end{equation}
\begin{figure}[t]
\centering
\includegraphics[width=0.75\textwidth]{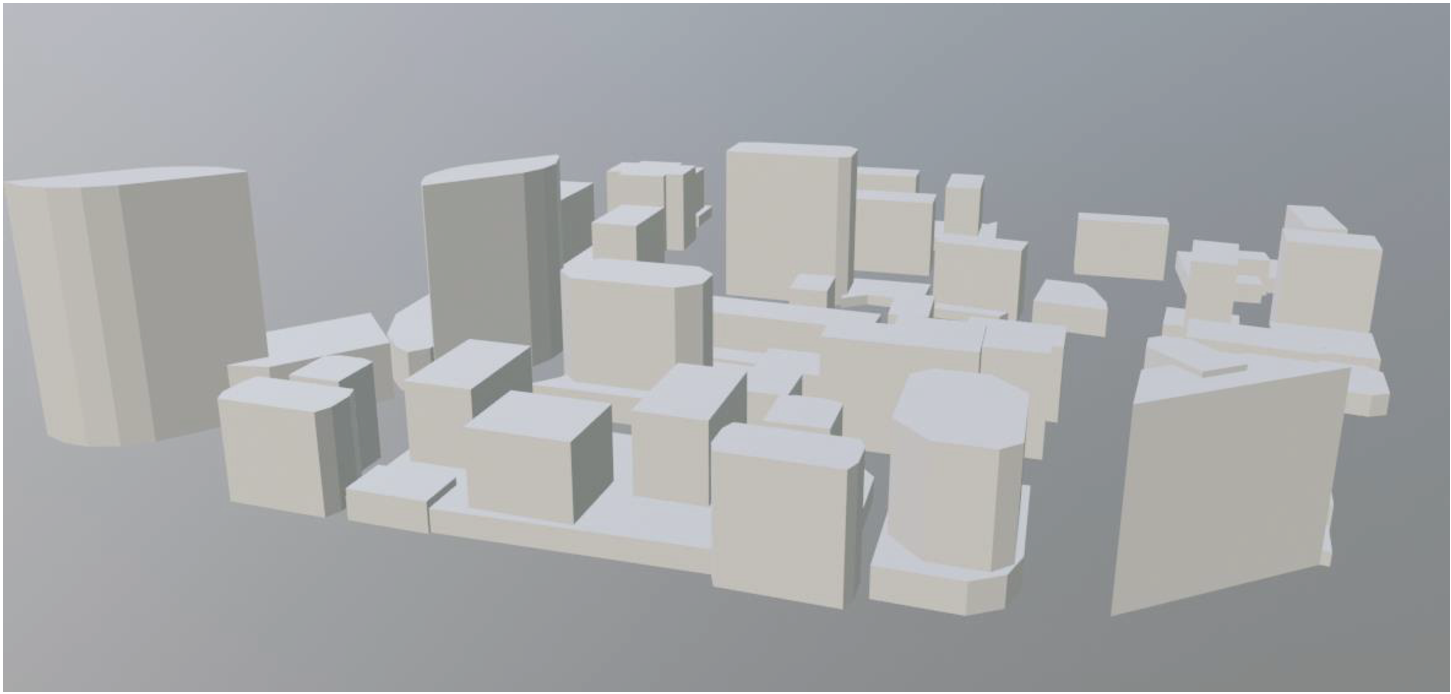}
\caption{The Rosslyn city model in Wireless Insite.}
\end{figure}
\subsection{Simulation Setup}
To evaluate ULCCME and SICCME, we resort to Wireless Insite \cite{tdma}, a ray tracing software\footnote{The ray tracing software has been adopted in many other research \cite{wang1}-\cite{Ahmed3}.} for channel generation to ensure a reasonable correlation between the channel and the environmental parameters, like user location, user trajectory, and the scene image, etc. The simulation steps are illustrated as follows:
\subsubsection{Channel Generation}
\begin{figure}[t]
	\begin{minipage}[t]{0.5\linewidth}
		\centering
	\includegraphics[width=78mm]{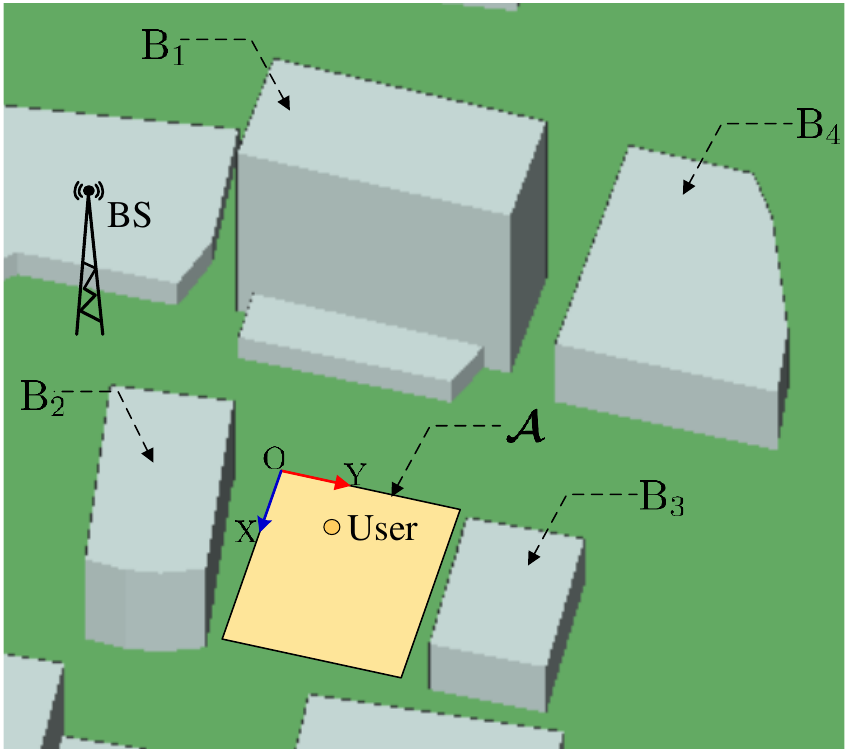}
		\caption{The simulated BS coverage area $\bm{\mathcal{A}}$.}
	\end{minipage}
	\hspace{1ex}
	\begin{minipage}[t]{0.5\linewidth}
		\centering
			\includegraphics[width=74mm]{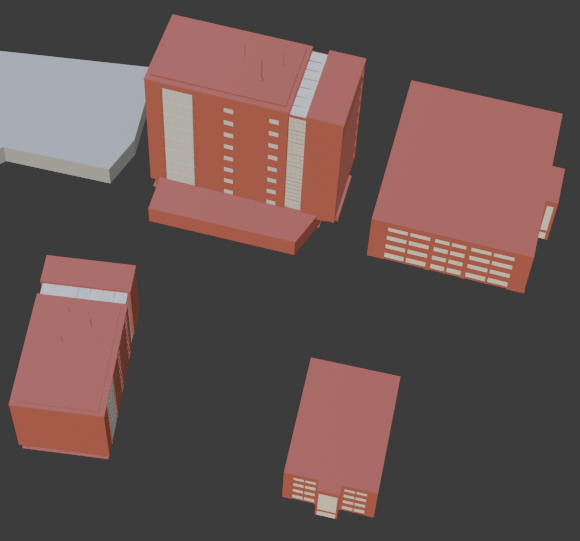}
		\caption{The city model with the rendered buildings.}
	\end{minipage}
\end{figure}
We adopt the Rosslyn city model in Wireless Insite as the 3D scene model of the BS's surrounding environment, as shown in Fig.~8. The BS coverage area $\bm{\mathcal{A}}$ is set as the selected rectangular area shown in Fig.~9. The size of $\bm{\mathcal{A}}$ is taken as $30\mathrm{m}\times 30\mathrm{m}$, and the coordinates of four vertices of $\bm{\mathcal{A}}$ are $(100, 160, 1.5)\mathrm{m}$, $(130, 160, 1.5)\mathrm{m}$, $(100, 190, 1.5)\mathrm{m}$ and $(130, 190, 1.5)\mathrm{m}$ respectively. The coordinates of BS is set to be $(75, 130, 10)\mathrm{m}$. For user at any location in $\bm{\mathcal{A}}$, Wireless Insite could produce  all parameters involved in the corresponding channel (2). Although the size of the selected region $\bm{\mathcal{A}}$ may be smaller than that of the actual BS coverage area, we consider the whole large BS coverage area can be divided into multiple small regions, and the DNN can be trained for each region to obtain the better learning accuracy than that of training a DNN for the whole BS coverage area, since the learning task of estimating covariance for the whole coverage area can be significantly more difficult.
\subsubsection{Scene Image Generation}
Though the Rosslyn city model in Wireless Insite is set to be the BS's surrounding environment, the original 3D model file does not have necessary texture information, and the color information of buildings can not  be reflected. The missing of texture information will cause serious lack of the visual distinction between the buildings in the original city model and make the scene images not realistic. We then use blender \cite{blender}, a 3D modeling software, to approximately replace the original buildings in the Rosslyn city model with the buildings rendered with proper texture. To reduce the repetition of the texture, only some buildings around the BS coverage area will be replaced by the rendered buildings. For example, as shown in Fig.~10, the buildings $\mathrm{B}_1,\mathrm{B}_2,\mathrm{B}_3,\mathrm{B}_4$ are selected to be replaced by the rendered buildings. Then, in the blender, the images to be taken by users can be simulated by the strategy in Section IV.

\subsubsection{Trajectory Generation}
\begin{figure}
  \centering
\subfigure[]{
\begin{minipage}[t]{0.5\linewidth}
\centering
\includegraphics[width=55 mm]{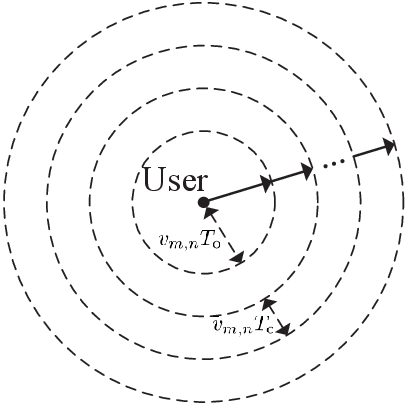}
\end{minipage}%
}%
\subfigure[]{
\begin{minipage}[t]{0.5\linewidth}
\centering
\includegraphics[width=55 mm]{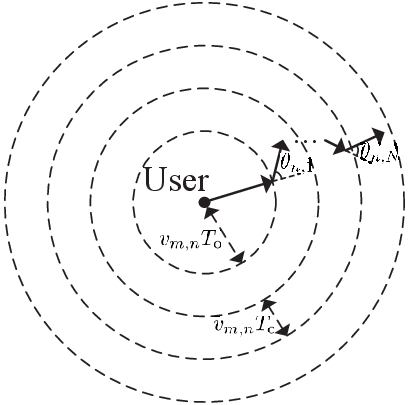}
\end{minipage}
}
\caption{(a) The $m$th simulated user trajectory with the constant direction pattern for the $n$th COCT. (b) The $m$th simulated user trajectory with the dynamic direction pattern for the $n$th COCT.}
\end{figure}
As shown in Section III, the user speed is assumed constant during one COCT. The user moving trajectory will affect the actual PDF of $\bm{x}_{\mathrm{u}}(T_{k,q}^{\mathrm{C}})$, as well as the performance of the proposed covariance estimation methods. Here, we consider the constant trajectory and dynamic trajectory of user to evaluate the channel estimation performance of ULCCME and SICCME, as shown in Fig.~11.

Specifically, we simulate $N_{\mathrm{tra}}$ trajectories for the user. Each simulated trajectory contains $N_{\mathrm{co}}$ COCTs. For the constant trajectory case, the user's moving direction does not change during one COCT. Hence, the $n$th COCT of the $m$th trajectory of the user, denoted as $\bm{x}_{\mathrm{u},m}(t)$, $t\in[T_{n}^{\mathrm{S}},T_{n+1}^{\mathrm{S}}]$, will be a straight line. The speed $v_{m,n}$ and the trajectory $\bm{x}_{\mathrm{u},m}(t)$, $t\in[T_{n}^{\mathrm{S}},T_{n+1}^{\mathrm{S}}]$ are generated from $[v_{\mathrm{L}},v_{\mathrm{U}}] \mathrm{m/s}$ and $[0,2\pi]$ through independent uniform sampling respectively, for any $m$ and $n$.

For the dynamic trajectory case, the user's moving direction will change at the beginning of each CCT in the $n$th COCT of the $m$th trajectory, $\forall m$ and $n$. The moving direction of the trajectory $\bm{x}_{\mathrm{u},m}(t)$, $t\in[T_{n}^{\mathrm{S}},T_{n,1}^{\mathrm{C}}]$ is generated from $[0,2\pi]$ by independent uniform distribution, and then the moving direction will change at $t=T_{n,q}^{\mathrm{C}}$, $q=1,2,\cdots,N$. The angle change $\theta_{n,q}$ of user moving direction at $t=T_{n,q}^{\mathrm{C}}$ obeys the truncated Gaussian distribution over $(-\pi,\pi)$ with $\mathcal{N}(0,\sigma_{\mathrm{a}}^2)$. Hence, the trajectory $\bm{x}_{\mathrm{u},m}(t)$, $t\in[T_{n}^{\mathrm{S}},T_{n+1}^{\mathrm{S}}]$, will be a piece-wise linear line. The speed $v_{m,n}$ of $\bm{x}_{\mathrm{u},m}(t)$, $t\in[T_{n}^{\mathrm{S}},T_{n+1}^{\mathrm{S}}]$ is also generated from $[v_{\mathrm{L}},v_{\mathrm{U}}] \mathrm{m/s}$ through independent uniform sampling, for any $m$ and $n$.
\subsubsection{Training Dataset Generation}
To generate the training dataset of LCNET, we select numerous user locations from $\bm{\mathcal{A}}$ with the uniform distribution. For each selected user location, we further choose numerous values from $[v_{\mathrm{L}}, v_{\mathrm{U}}] \mathrm{m/s}$ as the user speed with the uniform distribution. Then, for each user location and user speed, the output of LCNET is the vector shown in (13) that is constructed by the covariance calculated through (10), (11) and (12).

Note that the integral in (11) cannot be calculated accurately, as the channels of infinite location points need to be acquired. Hence, for a selected user location $\bm{x}_{\mathrm{s}}$ and a user moving speed $v_{\mathrm{s}}$, we only choose finite points from the estimated user moving region $\hat{\bm{C}}(\bm{x}_{\mathrm{s}},v_{\mathrm{s}}T_N)$ to approximate $\bm{\psi}(\bm{x}_{\mathrm{s}},v_{\mathrm{s}})$ in (11). Specifically, let us
 express $\bm{\psi}(\bm{x}_{\mathrm{s}},v_{\mathrm{s}})$ as
\begin{equation}
\begin{aligned}
&\bm{\psi}(\bm{x}_{\mathrm{s}},v_{\mathrm{s}})=\\
&\int_{0}^{2\pi}\int_{0}^{v_{\mathrm{s}}T_N}w(\rho,\theta) \bm{M}[\bm{c}(\bm{x}_{\mathrm{s}},\rho,\theta)]\bm{M}^{\mathrm{H}}[\bm{c}(\bm{x}_{\mathrm{s}},\rho,\theta)]\rho\mathrm{d}\rho\mathrm{d}\theta,
\end{aligned}
\end{equation}
where
\begin{equation}
\begin{aligned}
w(\rho,\theta)&=\frac{1}{N}\sum_{q=1}^{N}f_q(\rho,\theta),\\
f_q(\rho,\theta)&=\left\{\begin{array}{l}
                 \frac{1}{\pi v_{\mathrm{s}}^2T_q^2},\ \rho\leq v_{\mathrm{s}}T_q\\
                 0,\ \mathrm{otherwise}
               \end{array}\right.,
\end{aligned}
\end{equation}
where $w(\rho,\theta)$ is a probability density function of $\bm{x}=\bm{c}(\bm{x}_{\mathrm{s}},\rho,\theta)$ over $\hat{\bm{C}}(\bm{x}_{\mathrm{s}},v_{\mathrm{s}}T_N)$. When only finite points are chosen from $\hat{\bm{C}}(\bm{x}_{\mathrm{s}},v_{\mathrm{s}}T_N)$ to form the set $\bm{\mathcal{G}}$, the discrete probability distribution corresponding to these sampled points is given by
\begin{equation}
\mathrm{P}\{\bm{x}=\bm{x}^{\mathrm{s}}\}=\frac{w(||\bm{x}^{\mathrm{s}}-\bm{x}_{\mathrm{s}}||_2,\theta)}{\sum_{\hat{\bm{x}}\in\bm{\mathcal{G}}}w(||\bm{x}^{\mathrm{s}}-\bm{x}_{\mathrm{s}}||_2,\theta)},\ \forall \bm{x}^{\mathrm{s}}\in \bm{\mathcal{G}},\forall \theta,
\end{equation}
and can be used as an approximate estimate of $w(\rho,\theta)$. Thus, $\bm{\psi}(\bm{x}_{\mathrm{s}},v_{\mathrm{s}})$ can be approximately calculated as
\begin{equation}
\hat{\bm{\psi}}(\bm{x}_{\mathrm{s}},v_{\mathrm{s}})=\sum_{\bm{x}^{\mathrm{s}}\in\bm{\mathcal{G}}}\mathrm{P}\{\bm{x}=\bm{x}^{\mathrm{s}}\}\bm{M}(\bm{x}^{\mathrm{s}})\bm{M}^{\mathrm{H}}(\bm{x}^{\mathrm{s}}).
\end{equation}

\begin{figure*}[t]
\centering
\includegraphics[width=1\textwidth]{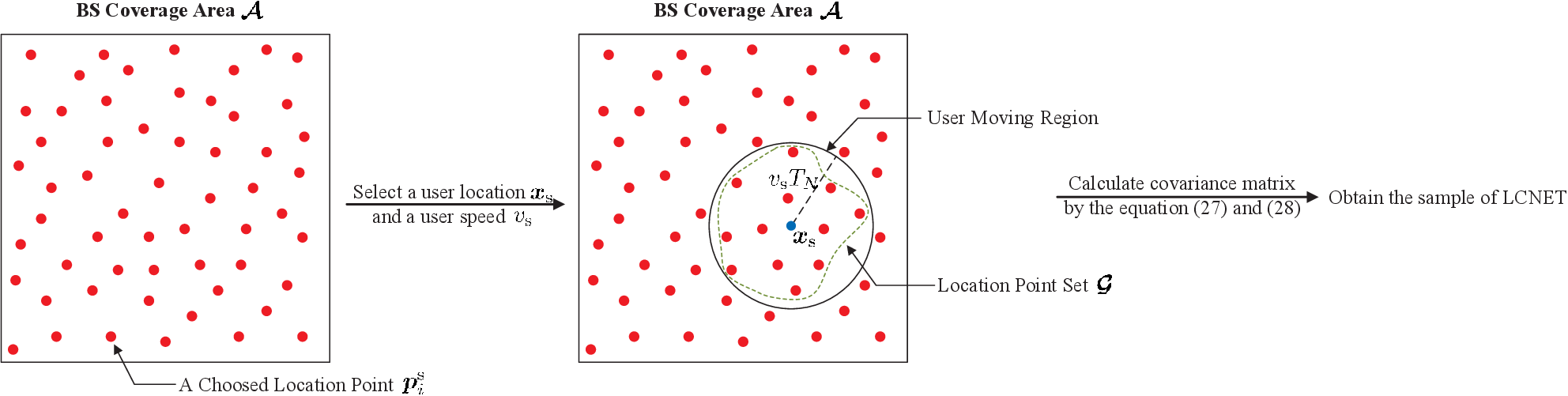}
\caption{The steps to construct the sample of LCNET.}
\end{figure*}

In the simulation, as the Fig.~12 shows, we uniformly choose $N_{\mathrm{p}}\gg 1$ points $\bm{p}^{\mathrm{s}}_1,\bm{p}^{\mathrm{s}}_2,\cdots,\bm{p}^{\mathrm{s}}_{N_{\mathrm{p}}}$ from $\bm{\mathcal{A}}$ and generate the channels of all these $N_{\mathrm{p}}$ points from Wireless Insite. Then, we select the points, which are contained in $\hat{\bm{C}}(\bm{x}_{\mathrm{s}},v_{\mathrm{s}}T_N)$, from $\bm{p}^{\mathrm{s}}_1,\bm{p}^{\mathrm{s}}_2,\cdots,\bm{p}^{\mathrm{s}}_{N_{\mathrm{p}}}$ to form the set $\bm{\mathcal{G}}$ and calculate (28). In practice, we can place the user, such as the intelligent vehicle, at many location points evenly throughout the BS coverage area for estimating the channels and recording the location coordinates of the corresponding channel. Thus, we can collect the channels and location coordinates of the necessary $N_{\mathrm{p}}$ points in $\bm{\mathcal{A}}$ to construct the samples of the LCNET. Although there is no doubt that the estimation error of the channel and location will exist in actual measurement, the effect of the estimation error can be very slight, since the channels can be estimated in high $\mathrm{SNR}$ conditions and the positioning accuracy of the current high-accuracy positioning techniques can even reach centimeter level with the development of the autonomous driving \cite{Kuutti}. It is worth mentioning that the coordinate space (CS) shown in Fig.~9 is adopted to determine the coordinates of the user location, where the coordinate space's origin is set as $\bm{\mathcal{A}}$'s vertex and $\bm{\mathcal{A}}$ is in the first quadrant of the coordinate space. The input feature of user location is the user plane coordinates $[\bm{x}]_{1:2}$ under the CS.

To train LENET, we also select numerous user locations from $\bm{\mathcal{A}}$ with uniform distribution and generate the corresponding channels of these selected locations to produce the input feature of LENET. The plane coordinates of these selected locations are set as the output label of LENET. By utilizing the corresponding channels $\bm{h}_{\mathrm{T},i}$, $i=1,2,\cdots,W_{\mathrm{t}}$, of these selected locations, we can approximately estimate $\zeta$ as $\zeta=\frac{1}{W_{\mathrm{t}}}\sum_{i=1}^{W_{\mathrm{t}}}||\bm{h}_{\mathrm{T},i}||_2$.

For each training sample, four scene images are taken at the corresponding user location by blender to generate the input feature $\hat{\bm{F}}$ of ICNET, as shown in Section V, and the output of ICNET is the same as the label of this sample. Then, the dataset of ICNET and LCNET will have the same number of samples. The image resolution in our simulations is taken as  $270\times480$.

Note that, though the target output covariance is $\bm{R}_i$ for the $i$th training sample of LCNET or ICNET, the practically adopted output label of the $i$th training sample is the vector (13) constructed by the normalized covariance $\frac{N_{\mathrm{B}}S\bm{R}_i}{\sum_{i=1}^{S}\mathrm{tr}(\bm{R}_i)}$ to improve the learning performance of the DNNs, where $S$ is the number of training samples. Thus, all the covariance output from LCNET or ICNET should be scaled by the coefficient $\frac{\sum_{i=1}^{S}\mathrm{tr}(\bm{R}_i)}{N_{\mathrm{B}}S}$ to recover the original CCM.
\subsubsection{Simulation Parameters}
\begin{table}[t]
\centering
\caption{Critical Parameters of Wireless Insite for Ray Tracing}
\begin{tabular}{cccc}
\toprule
Parameter& Value\\
\midrule
Carrier Frequency& 2.4 $\mathrm{GHz}$\\
Propagation Model& X3D\\
Building Material& Concrete\\
Maximum Number of Reflections& 6\\
Maximum Number of Diffractions& 1\\
Maximum Paths Per Receiver Point& 50\\
\bottomrule
\end{tabular}
\end{table}
The dimensions of the array at BS are taken as $N_{\mathrm{ele}}^{\mathrm{B}}=3$ and $=N_{\mathrm{az}}^{\mathrm{B}}=4$, and the critical parameters of Wireless Insite to generate channels are illustrated in TABLE.~I; $T_\mathrm{o}=T_\mathrm{u}+T_\mathrm{e}$ and $T_\mathrm{c}$ are both set to be 0.005s; $N_{\mathrm{tra}}=20$ trajectories are generated, and each trajectory contains $N_{\mathrm{co}}=10$ COCTs; $N$ and $N_{\mathrm{p}}$ are set to be 50 and 62500 respectively. For each generated trajectory, the user speed is uniformly selected from $[v_{\mathrm{L}},v_{\mathrm{U}}]=[2,10]\ \mathrm{m/s}$; $Q$ and $M_{\mathrm{p}}$ are set to be 20 and 60, respectively; $\tilde{\sigma}^2$ is set to satisfy $\frac{N_{\mathrm{B}}\tilde{\sigma}^2}{E\{||\bm{h}||_2^2\}}=10^{-2}$, where $E\{||\bm{h}||_2^2\}$ is approximately  $\frac{1}{W_{\mathrm{t}}}\sum_{i=1}^{W_{\mathrm{t}}}||\bm{h}_{\mathrm{T},i}||_2^2$.

For the LCNET, the node number of each layer of the subnetwork corresponding to user location before the concatenate layer is set to be 50 and 100 respectively and the node number of each layer of the subnetwork for user speed is set to be 20 and 50 respectively. The node numbers of the fully connected layers that begin with the concatenate layer and end with the activation layer are set to be 150, 150, and 150 respectively. The node numbers of the fully connected layers that begin with the activation layer are set to be 200, 200, 150, 150 and 144 respectively. ICNET$_{\mathrm{U}}$ has four residual blocks (ResBlocks) with CBAM and three 2D convolutional layers and the last seven fully connected layers with 2000, 1000, 500, 200, 100, 50 and 2 nodes, respectively. The kernel size of convolutional layer of each ResBlock's spatial attention module are set to be (7,7), (5,5), (3,3) and (1,1), respectively. The parameters of the pooling layer, convolutional layers and ResBlocks of ICNET$_{\mathrm{U}}$ are shown in TABLE.~II. Moreover, batch normalization is applied after each convolutional layer of ICNET$_{\mathrm{U}}$ except the convolutional layers in spatial attention module. For the ICNET$_{\mathrm{B}}$, the node number of each layer of the subnetwork corresponding to user speed before the concatenate layer is set to be 20 and 50 respectively and the node number of each layer of the subnetwork for the image feature is set to be 50 and 100 respectively. The node numbers of the fully connected layers that begin with the concatenate layer are set to be 200, 200, 150, 150 and 144 respectively. The node number of each layer of LENET is set to be 50, 100, 200, 100, 50 and 2 respectively. The adopted activation function and optimizer for LCNET, LENET and ICNET are ReLU/PReLU and Adam respectively.
\begin{table}[t]
\centering
\caption{The Parameters of Convolutional/Pooling Layers of ICNET$_{\mathrm{U}}$}
\begin{tabular}{cccc}
\toprule
Layer Order& Kernel/Pool Size& Strides& Filters\\
\midrule
AvgPooling& (2, 2)& (2, 2)& None\\
ResBlock with CBAM& (3, 3)& (1, 1)& 12\\
Convolutional& (6, 5)& (2, 2)& 6\\
ResBlock with CBAM& (3, 3)& (1, 1)& 6\\
Convolutional& (2, 2)& (2, 2)& 8\\
ResBlock with CBAM& (3, 3)& (1, 1)& 8\\
Convolutional& (2, 2)& (2, 2)& 4\\
ResBlock with CBAM& (3, 3)& (1, 1)& 4\\
\bottomrule
\end{tabular}
\end{table}

The training and test set of LCNET and ICNET are constructed by 30000 and 1000 selected user locations respectively. For each selected user location in the training set, 40 speed values are selected from $[v_{\mathrm{L}}, v_{\mathrm{U}}] \mathrm{m/s}$. For each selected user location in the test set, 5 speed values are selected from $[v_{\mathrm{L}}, v_{\mathrm{U}}] \mathrm{m/s}$. The training set of LENET includes $W_{\mathrm{t}}=107500$ samples. The additional sample set of LENET utilized to obtain $\bm{m}_{\mathrm{n}}$ and $\bm{R}_{\mathrm{n}}$ includes $W=9000$ samples.

\subsection{Results and Discussions}
\begin{figure}[t]
\centering
\includegraphics[width=0.7\textwidth]{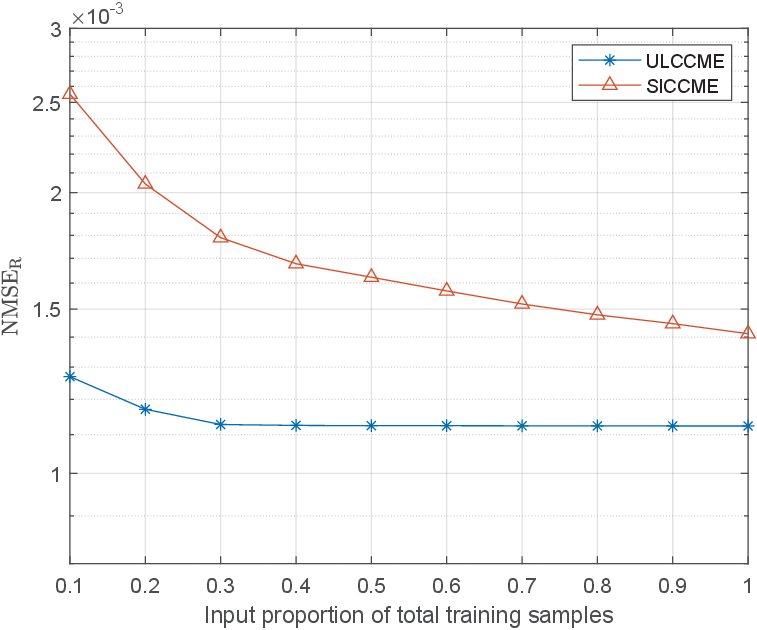}
\caption{$\mathrm{NMSE}_{\mathrm{R}}$ of ULCCME and SICCME with different input training set sizes.}
\end{figure}
We first analyze the learning accuracies of ULCCME and SICCME under different training set sizes in Fig.~13. With the increase of input proportion of training samples, the $\mathrm{NMSE}_{\mathrm{R}}$ of ULCCME and SICCME all decrease while ULCCME is better than SICCME. This is not unexpected since ULCCME relies on more accurate location information than SICCME. Nevertheless, the difference is not severe, which  demonstrates the effectiveness of purely relying on scene images.

\begin{figure}[t]
	\begin{minipage}[t]{0.5\linewidth}
		\centering
	\includegraphics[width=83mm]{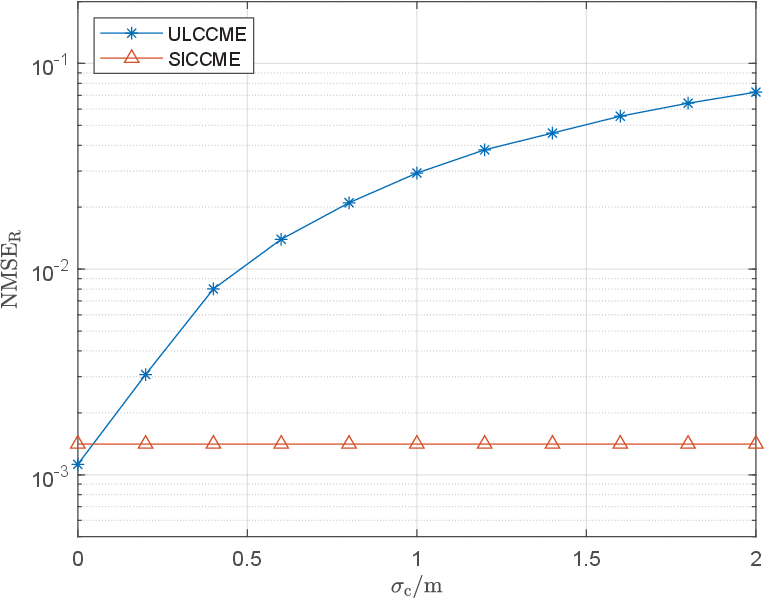}
		\caption{$\mathrm{NMSE}_{\mathrm{R}}$ of ULCCME and SICCME under different $\sigma_{\mathrm{c}}$.}
	\end{minipage}
	\hspace{1ex}
	\begin{minipage}[t]{0.5\linewidth}
		\centering
			\includegraphics[width=83mm]{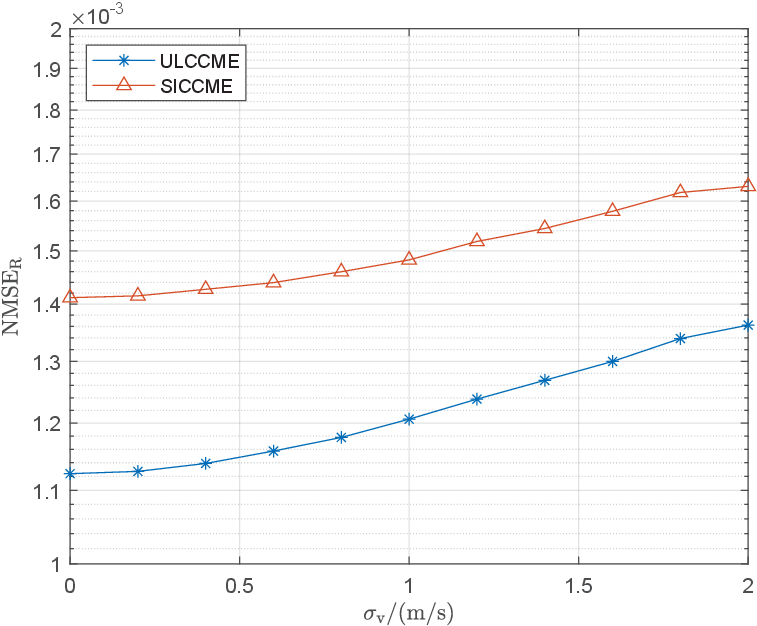}
		\caption{$\mathrm{NMSE}_{\mathrm{R}}$ of ULCCME and SICCME under different $\sigma_{\mathrm{v}}$.}
	\end{minipage}
\end{figure}

We further consider the influence of the estimation error of the uploaded user location and user speed on the CCM estimation accuracy. We assume the estimation error of user location and user speed follows the Gaussian distribution $\mathcal{N}(0,\sigma_{\mathrm{c}}^2)$ and $\mathcal{N}(0,\sigma_{\mathrm{v}}^2)$, respectively. If the coordinates of the noisy location are out of the area $\bm{\mathcal{A}}$, we will correct it by the following way: Denote the coordinate of any dimension of the uploaded location as $x$, and define the minimum/maximum bound coordinate of $\bm{\mathcal{A}}$ at the dimension corresponding to $x$ as $x_{\mathrm{L}}$/$\mathrm{x}_{\mathrm{U}}$. Since the actual user location must be in $\bm{\mathcal{A}}$, the coordinate value $x$ will be corrected as $\max(\min(x,x_{\mathrm{U}}),x_{\mathrm{L}})$. We utilize the same principle to correct the uploaded user speed if the value of the uploaded speed is out of the interval $[v_{\mathrm{L}},v_{\mathrm{U}}]\mathrm{m/s}$.

We then demonstrate the $\mathrm{NMSE}_{\mathrm{R}}$ of ULCCME and SICCME under different location noise variance $\sigma^2_{\mathrm{c}}$ and speed noise variance $\sigma^2_{\mathrm{v}}$ in Fig.~14. It is seen that the $\mathrm{NMSE}_{\mathrm{R}}$ of SICCME remains constant for different $\sigma_{\mathrm{c}}$. The reason is that SICCME is related with environment image and will not be affected by the location noise. It is also seen that the $\mathrm{NMSE}_{\mathrm{R}}$ of ULCCME deteriorates with the increasing of $\sigma_{\mathrm{c}}$, and will be worse than that of SICCME once $\sigma_{\mathrm{c}}>0.1\mathrm{m}$. Hence, under the scenarios with serious location noise, the SICCME will achieve better performance than the ULCCME method. As shown in Fig.~15, the $\mathrm{NMSE}_{\mathrm{R}}$ of ULCCME and SICCME under different $\sigma_{\mathrm{v}}$ are plotted. It is seen that the $\mathrm{NMSE}_{\mathrm{R}}$ of ULCCME and SICCME all deteriorates as $\sigma_{\mathrm{v}}$ increases. It is also seen that the performance of ULCCME and SICCME is more robust to the speed noise than to the location noise. For example, when $\sigma_{\mathrm{v}}=2\mathrm{m/s}$, the $\mathrm{NMSE}_{\mathrm{R}}$ of ULCCME and SICCME only increase by $21.2\%$ and $15.5\%$, respectively, though the $\mathrm{NMSE}_{\mathrm{R}}$ of ULCCME increases by $6354.13\%$ when $\sigma_{\mathrm{c}}=2\mathrm{m}$. Therefore, in following simulations, we will mainly demonstrate the performance of ULCCME and SICCME caused by the location noise.

\begin{figure}[t]
\centering
\includegraphics[width=0.7\textwidth]{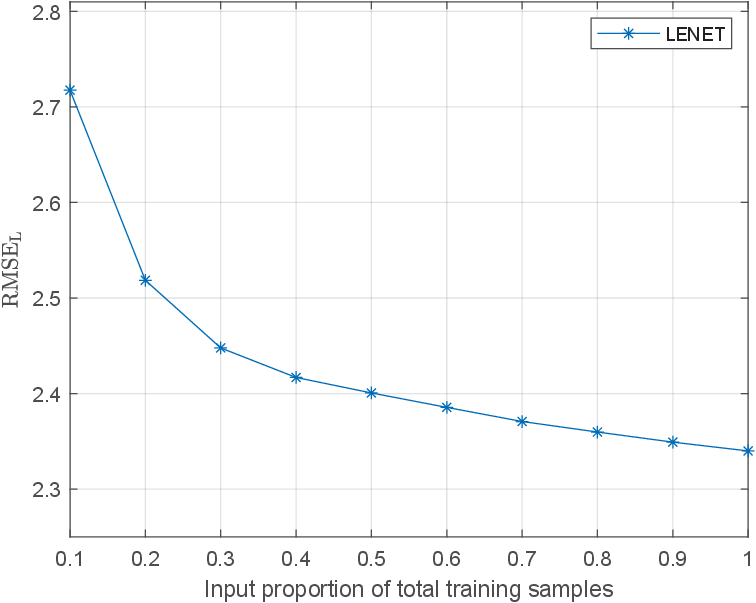}
\caption{$\mathrm{RMSE}_{\mathrm{L}}$ of LENET with different input training set sizes.}
\end{figure}

As the performance of the proposed location denoising method depends on the location estimation accuracy of LENET, we plot the $\mathrm{RMSE}_{\mathrm{L}}$ of LENET under different input proportions of the training set in Fig.~16. It is seen that the optimal location estimation performance of LENET can achieve $\mathrm{RMSE}_{\mathrm{L}}=2.34\mathrm{m}$ if the overall training set is used. It is also seen that the decreasing of $\mathrm{RMSE}_{\mathrm{L}}$ slows down when more training samples are used. Thus, only half of the dataset is needed to achieve $\mathrm{RMSE}_{\mathrm{L}}=2.40\mathrm{m}$ that approaches the optimal location estimation performance.

\begin{figure}[t]
	\begin{minipage}[t]{0.5\linewidth}
		\centering
	\includegraphics[width=83mm]{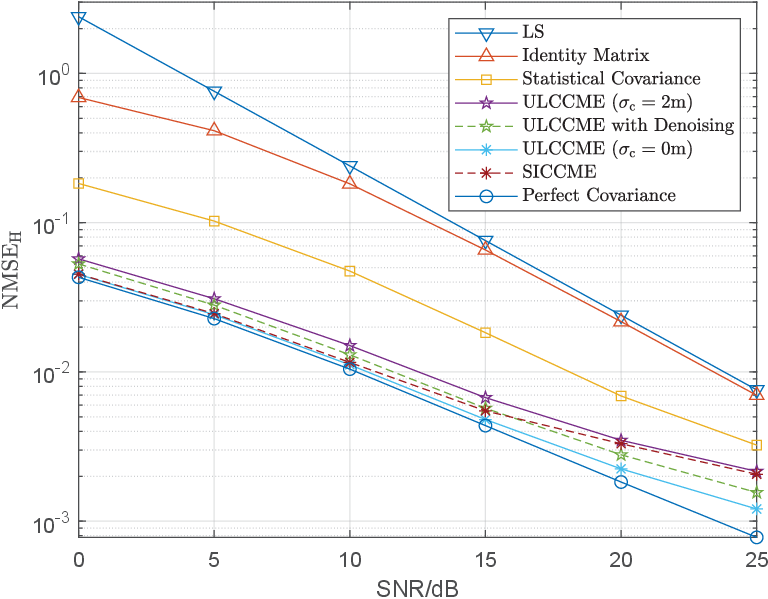}
		\caption{$\mathrm{NMSE}_{\mathrm{H}}$ of different covariance estimation methods for the constant trajectory case.}
	\end{minipage}
	\hspace{1ex}
	\begin{minipage}[t]{0.5\linewidth}
		\centering
			\includegraphics[width=83mm]{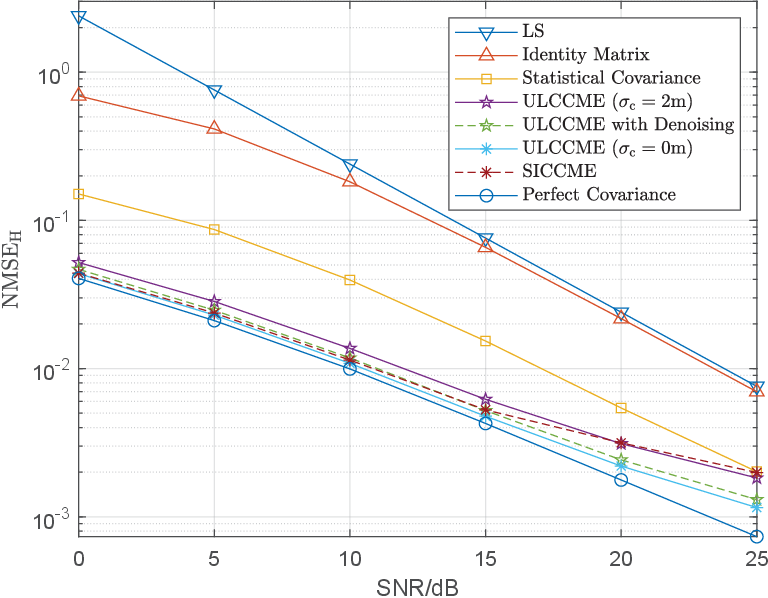}
		\caption{$\mathrm{NMSE}_{\mathrm{H}}$ of different covariance estimation methods for the dynamic trajectory case.}
	\end{minipage}
\end{figure}

We then compare the channel estimation performance with the CCM obtained from ULCCME, SICCME, as well as two traditional methods \cite{Biguesh} and the statistical method based on historical estimated channels, in Fig.~17 and Fig.~18 respectively. The first traditional method is the least square (LS) channel estimator, while the second traditional method utilizes the normalized identity matrix $\frac{\sum_{i=1}^{S}\mathrm{tr}(\bm{R}_i)}{N_{\mathrm{B}}S}\bm{I}_{N_\mathrm{B}}$ as the CCM to design the pilot signal matrix as well as to perform LMMSE channel estimation. The statistical method utilize $\frac{\sum_{i=1}^{S}\mathrm{tr}(\bm{R}_i)}{N_{\mathrm{B}}S}\bm{I}_{N_\mathrm{B}}$ as the CCM for the first COCT of each trajectories. For the next COCT, the statistical method utilize the $N$ channels estimated in the previous COCT to calculate a statistical covariance for channel estimation. It is worth noting that the statistical method will need the user to accurately feed back the estimated channels of $N$ CCTs or the obtained statistical covariance during each COCT, which means huger feedback cost than other compared methods. We also present the channel estimation performance of ULCCME with/without the location denoising and SICCME method. Moreover, the channel estimation performance achieved by the perfect CCM $\frac{1}{N}\sum_{q=1}^{N}\bm{M}[\bm{x}_{\mathrm{u},m}(T^{\mathrm{C}}_{n,q})]\bm{M}^{\mathrm{H}}[\bm{x}_{\mathrm{u},m}(T^{\mathrm{C}}_{n,q})]$ is also provided as benchmark. The $\mathrm{SNR}$ is defined as
\begin{equation}
\mathrm{SNR}=\frac{P}{M_{\mathrm{p}}\sigma^2 N_{\mathrm{tra}}N_{\mathrm{co}}N}\sum_{m=1}^{N_{\mathrm{tra}}}\sum_{n=1}^{N_{\mathrm{co}}}\sum_{q=1}^{N}||\bm{M}[\bm{x}_{\mathrm{u},m}(T^{\mathrm{C}}_{n,q})]||_2^2.
\end{equation}
As shown in Fig.~17 and Fig.~18, the $\mathrm{NMSE}_{\mathrm{H}}$ of the ULCCME and SICCME method is better than that of the two traditional methods and the statistical method, even when there are location noise with variance $\sigma_{\mathrm{c}}=2\mathrm{m}$. This indicates that though the accurate probability distribution of $\bm{x}_{\mathrm{u}}(T_{k,q}^{\mathrm{C}})$ is replaced by uniform distribution to simplify the analysis, the proposed CCM estimation methods can still outperform the compared methods under different trajectory case. It is interesting that the statistical method achieves worse channel estimation accuracy under the constant trajectory case than the dynamic trajectory case. This is due to that for the constant trajectory case, the invariance of user moving direction makes the average overlapping area between the user moving region during the previous COCT and the current COCT be smaller than for the dynamic trajectory case. However, the proposed CCM estimation methods can achieve similar and near optimal performance under the two trajectory cases. Meanwhile, although the optimal $\mathrm{RMSE}_{\mathrm{L}}=2.34 \mathrm{m}$, i.e., the standard deviation of location estimation error of LENET, is worse than that with the standard deviation $\sigma_{\mathrm{c}}=2\mathrm{m}$ of the location noise, the proposed location denoising method can still effectively improve the channel estimation performance of ULCCME method with the aid of LENET. It is also seen that when there is no location noise, the channel estimation performance of the SICCME method is worse than that of the ULCCME method, because the ULCCME can achieve better CCM estimation than SICCME. However, for the case with location noise $\sigma_{\mathrm{c}}=2\mathrm{m}$, the channel estimation performance achieved by ULCCME will be worse than that of SICCME, when the SNR is lower than $20\mathrm{dB}$. Moreover, the channel estimation with the proposed ULCCME and SICCME methods can approach the optimal channel estimation that is achieved by the perfect covariances under relatively lower SNR.

\begin{figure}[t]
	\begin{minipage}[t]{0.5\linewidth}
		\centering
	\includegraphics[width=83mm]{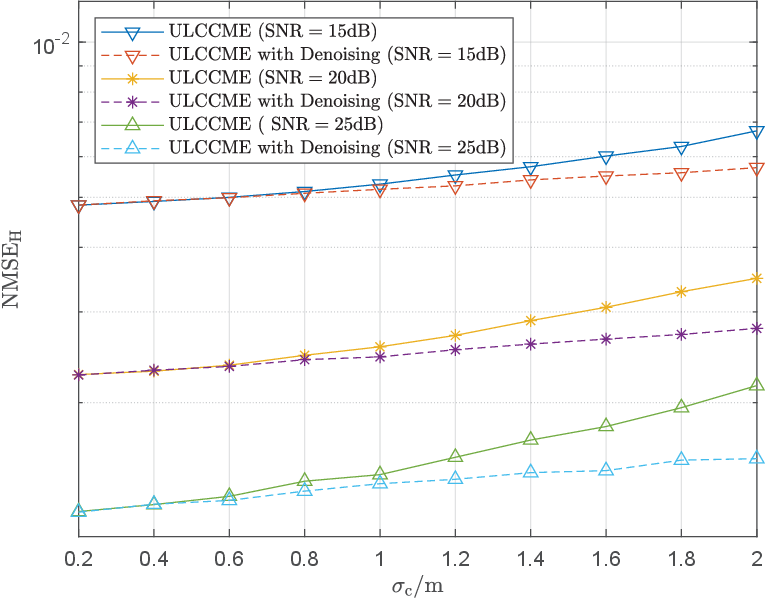}
		\caption{$\mathrm{NMSE}_{\mathrm{H}}$ of location based method with/without location denoising under different $\sigma_{\mathrm{c}}$ for the constant trajectory case.}
	\end{minipage}
	\hspace{1ex}
	\begin{minipage}[t]{0.5\linewidth}
		\centering
			\includegraphics[width=83mm]{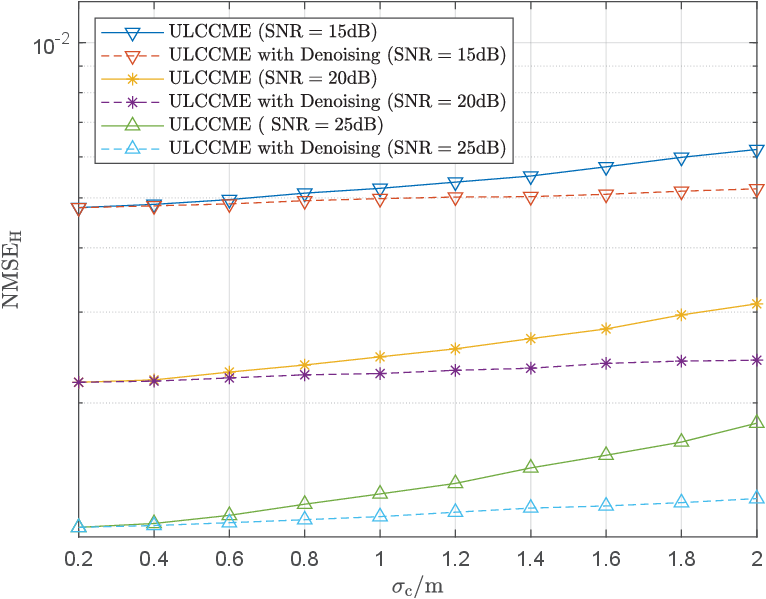}
		\caption{$\mathrm{NMSE}_{\mathrm{H}}$ of location based method with/without location denoising under different $\sigma_{\mathrm{c}}$ for the dynamic trajectory case.}
	\end{minipage}
\end{figure}

To demonstrate the effectiveness of the location denoising method, we plot the channel estimation performance of the ULCCME method with/without the location denoising under different $\sigma_{\mathrm{c}}$ in Fig.~19 and Fig.~20, respectively. It is seen that the performance improvement by location denoising is greater with larger location noise and  lower $\mathrm{SNR}$, which indicates the proposed location denoising method is more suitable for the scenario with serious positioning noise and low $\mathrm{SNR}$.

\section{Conclusion}
We have proposed two deep learning based methods for CCM estimation by utilizing the environmental information, e.g., user location and scene images, that can outperform the traditional CCM estimation method that relies on historical estimated channels. A location denoising method is also designed to improve the robustness of ULCCME  by jointly analyzing the estimation error of the trained DNN and position error of user. In fact, if the user motion contains certain regularity, then a more accurate probability density for the user location can be derived to further improve the effectiveness of ULCCME. Simulation results show that the proposed ULCCME and SICCME are effective and can achieve better channel estimation than the traditional ones. Moreover, the proposed SICCME method has significant performance gain compared with the ULCCME method when the location noise is larger or the user location is unavailable, which also motivates us to further explore the scene image information into the field of wireless communications.
\balance

\end{document}